\documentclass{article}
 \pdfoutput=1
\usepackage{graphicx}  % standard LaTeX graphics tool;
\usepackage{amsmath}   % For getting proper math eqns
\usepackage{amssymb}   % Used for getting various symbols eg. gtrsim, lesssim etc.
\usepackage[export]{adjustbox}
\usepackage{bm} % For bold math (esp of lower greek letters)
\usepackage{dcolumn}% Align table columns on decimal point
\usepackage{color}
\usepackage{mathrsfs}
\usepackage{amsfonts}
\usepackage{varioref}

\usepackage{physics}
\usepackage[vcentermath]{youngtab}
\usepackage{wrapfig}

\RequirePackage[colorlinks,citecolor=blue,urlcolor=magenta,linkcolor=blue]{hyperref}

\def\Tr{{\rm Tr}}

\labelformat{section}{Section #1} 
\labelformat{subsection}{Section #1} 
\labelformat{subsubsection}{Section #1}
\labelformat{subsubsubsection}{Section #1}
\labelformat{equation}{Eq.~(#1)} 
\labelformat{figure}{Fig.~#1} 
\labelformat{subfigure}{Fig.~\thefigure#1} 
\labelformat{table}{Tab.~#1} 
\labelformat{appendix}{Appendix #1}
\title{\bf IR dynamics and entanglement entropy}
\author{\bf Theodore N Tomaras$^{1}$\footnote{tomaras@physics.uoc.gr}
\bf ~and~Nicolaos Toumbas$^{2}$\footnote{nick@ucy.ac.cy}\\
$^{1}$\small{ITCP and Department of Physics, University of Crete, 700 13 Heraklion, Greece}\\
$^{2}$\small{Department of Physics, University of Cyprus, Nicosia 1678, Cyprus} }

\date{ }  %% This command  will suppress printing the date. 
\begin{document}
  
\maketitle
\begin{abstract}
\noindent
We consider scattering of Faddeev-Kulish electrons in QED and study the entanglement between the hard and soft particles in the final
state at the perturbative level. The soft photon spectrum naturally splits into two parts: i) soft photons with energies less than a characteristic infrared scale $E_d$ present in the clouds accompanying the asymptotic charged particles, and ii) sufficiently low energy photons with energies greater than $E_d$, comprising the soft part of the emitted radiation. We construct the density matrix associated with tracing over the radiative soft photons and calculate the entanglement entropy perturbatively. We find that the entanglement entropy is free of any infrared divergences order by order in perturbation theory. On the other hand infrared divergences in the perturbative expansion for the entanglement entropy appear upon tracing over the entire spectrum of soft photons, including those in the clouds. To leading order the entanglement entropy is set by the square of the Fock basis amplitude for real single soft photon emission, which leads to a logarithmic infrared divergence when integrated over the photon momentum. We argue that the infrared divergences in the entanglement entropy (per particle flux per unit time) in this latter case persist to all orders in perturbation theory in the infinite volume limit.             
\end{abstract}
\vskip .5cm 
\noindent
%{\bf PACS : } \\
\noindent
{\bf keywords : Entanglement entropy, soft theorems, IR divergences} 
\bigskip
%%%%%%%%%%%%%%%%%%%%%%%%%%%%%%%%%%%%%%%%%%%%%%%%%%%%%%%%%%%%%%%%%%%%%%%%%%%%%%%%%%
\section{Introduction}\label{s1}
%%%%%%%%%%%%%%%%%%%%%%%%%%%%%%%%%%%%%%%%%%%%%%%%%%
\noindent
Symmetry renders purely hard scattering processes in QED and gravity impossible \cite{StromingerLectures,StromingerIRrevisited,StromingerBMS,HMPS,LPS,KMS,Banks1,SZ,Weinberg}. Rather the asymptotic particles must be accompanied by infinite clouds of soft photons or gravitons, in addition to the soft radiation emitted during the process. The hard and soft particles are highly correlated. As the resolution of particle detectors is limited, an infinite number of soft particles evades detection in a typical experiment. It is therefore important to understand the nature of the entanglement between the hard and soft degrees of freedom in the final state, and to quantify the information carried by the soft particles. A measure of this information is provided by the entanglement entropy. In \cite{StromingerBHinfo} it was argued that soft quanta emitted during the process of formation/evaporation of a four-dimensional black hole could play an important role in the resolution of the black hole information paradox. See also \cite{Banks2} for related discussions as well as extensions to higher dimensions.

Indeed as shown in \cite{Carney1,Carney2,Carney3}, tracing over the soft particles in the final state can lead to decoherence, revealing strong entanglement between the hard and soft degrees of freedom. See also \cite{Gomez1, Gomez2}. In this paper we consider typical scattering processes in QED in order to study the reduced density matrix and calculate the entanglement entropy perturbatively. We focus on the example of electron-electron scattering to illustrate our results. To regulate the entanglement entropy, we discretize the system by putting the process in a large box of size $L$ and impose an infrared cutoff $\lambda$ of order $1/L$. At the end of the calculation, we take the continuum, $\lambda \to 0$ limit. We would like to investigate if infrared divergences in the entanglement entropy appear, and whether they cancel order by order in perturbation theory. We discuss both a Fock basis computation where we take a state of two bare electrons for the initial state and a proper asymptotic state where the electrons are ``dressed'' with infinite clouds of soft photons, in accordance with the Faddeev-Kulish construction \cite{FK,Chung}. Other pertinent work on entanglement after scattering includes \cite{Bala,Seki,Seki2,Grignani,Petruccione,Calucci,Asorey}.

First we trace over the entire soft part of the Hilbert space, comprised of photon states with total energy less than an infrared energy scale $E$ (smaller than the mass of the electron). This energy scale is set by the sensitivity of the detector. The reduced density matrix is an operator acting on the hard part of the Hilbert space, and (to all orders in perturbation theory) it exhibits decoherence in the continuum limit \cite{Carney2,Carney3}. We find logarithmic infrared divergences in the perturbative expansion for the entanglement entropy, for both the dressed and the Fock basis computations. In both cases and to leading order in perturbation theory, the entanglement entropy is proportional to the conventional Fock basis rate for the two initial electrons to scatter and emit at the same time a single soft photon with frequency in the range $\lambda<\omega_\gamma <E$. This rate diverges logarithmically in the continuum, $\lambda \to 0$ limit at tree level. For the Fock basis calculation the infrared divergence can be attributed to the soft part of the emitted radiation. For the case of Faddeev-Kulish electrons, the divergence can be traced in the overlap of the coherent states describing the soft photon clouds dressing the final state charged particles. Despite the fact that the Faddeev-Kulish $S$-matrix is infrared finite order by order in perturbation theory \cite{FK, Chung}, the dressing does not alleviate logarithmic divergences in the entanglement entropy at the perturbative level. In fact the leading perturbative entanglement entropy is a fraction of the maximal possible value, as set by the dimensionality of the subspace of single soft photon states. We argue that the structure of the singular part is universal, applicable to generic scattering processes, and show that the coefficient of the IR logarithmic singularity is related to the cusp anomalous dimension in QED. We also argue that infrared logarithmic divergences in the entanglement entropy (per particle flux per unit time) persist to all orders in the infinite volume limit.

On the other hand, the Faddeev-Kulish cross-section for the emission of soft photons of energy less than $E_d$, the scale characterizing the photons in the clouds, is suppressed (and likewise for gravitons) \cite{Choi2}. Thus we may distinguish between soft cloud photons and radiated ones in the final state. These observations motivate us to consider a second type of partial trace, over soft photons with frequencies in the range $E_d < \omega_\gamma < E$, comprising in the dressed case the soft part of the emitted radiation. This type of tracing was also advocated in \cite{Gomez1,Gomez2} in order to alleviate the amount of decoherence in the continuum limit. The reduced density matrix is now an operator acting on the space of asymptotic states. Both the diagonal and off diagonal elements are given in terms of Faddeev-Kulish amplitudes, which are free of any infrared divergences in the continuum $\lambda \to 0$ limit (order by order in perturbation theory). The entanglement entropy is finite order by order in perturbation theory. The leading entanglement entropy can be expressed in terms of the Fock basis rate for the emission of a single soft photon with frequency in the range $E_d<\omega_\gamma <E$. When the IR scales $E_d$ and $E$ are sufficiently small, this rate is proportional to the logarithm of the ratio of the infrared scales $E/E_d$, which remains finite in the $\lambda \to 0$ limit. The perturbative analysis is now valid and can be trusted in the continuum limit. In effect the dressing provides an infrared cutoff of order the cloud energy scale $E_d$, curing the singular behavior associated with the previous tracing. As $E \to E_d$, the entanglement entropy becomes very small, and therefore we conclude that a small amount of information is carried by the extra soft radiated photons. These results are consistent with estimates of the amount of decoherence obtained in \cite{Gomez2}.

The plan of the paper is as follows. In \ref{s2} we describe the various infrared energy scales and decompose the Hilbert space into soft and hard factors. We then review the Faddeev-Kulish construction of asymptotic states in QED and exhibit the finiteness of the $S$ matrix. The reader familiar with this construction may omit most material in this section. In \ref{s3} we describe the discretization of the system by replacing infinite space with a large box of finite size and impose an infrared cutoff. We construct dressed states for the discrete system reproducing the Faddeev-Kulish states in the continuum limit. We also explain how to compute various partial traces, which will be useful for the following calculations. In \ref{s4} we consider a two electron scattering process and construct the reduced density matrices after taking the partial tracings over the final state, as outlined above. Keeping the infrared cutoff $\lambda$ finite, we compute the Renyi entropies (for integer $m$) and the entanglement entropy to leading order in perturbation theory. For the dressed case, restricting the trace over the soft part of the emitted radiation, yields a finite entanglement entropy (per unit flux per unit time), free of any infrared divergences in the continuum, $\lambda \to 0$ limit. We summarize our results and discuss implications and open problems in \ref{s5}.

%%%%%%%%%%%%%%%%%%%%%%%%%%%%%%%%%%%%%%%%%%%%%%%%%%%%%%%%%%%%%%%%%%%%%%%%%%%%%%%%%%
\section{QED scattering, soft photons and entanglement}\label{s2}
%%%%%%%%%%%%%%%%%%%%%%%%%%%%%%%%%%%%%%%%%%%%%%%%%%
\noindent
Scattering processes in QED are constrained by an infinite set of conservation laws associated with large gauge transformations (LGT) \cite{StromingerLectures,StromingerIRrevisited,HMPS,LPS,KMS,Campiglia}. These are transformations that do not vanish at infinity, but instead approach angle dependent constants. Transitions between conventional Fock states, where only a finite number of photons are present in the initial and final states, fail to satisfy the conservation laws associated with LGT. As a result the corresponding $S$-matrix elements vanish \cite{StromingerIRrevisited}. An infinite number of soft photons must be present in the final state. The vanishing of the Fock basis transition amplitudes is more commonly attributed to the exponentiation of virtual infrared divergences, see e.g. \cite{Weinberg}, but it can also be understood as a consequence of symmetry.

On the other hand, the conventional Fock basis states do not diagonalize the asymptotic Hamiltonian, which includes the slowly decaying parts of the interaction Hamiltonian (written in the interaction picture). As shown by Faddeev and Kulish, physical asymptotic states can be constructed by dressing the Fock charged particle states with clouds of soft photons \cite{FK}. The $S$-matrix elements between these dressed states are nonvanishing and free of infrared divergences \cite{FK, Chung, Kibble1, Kibble2}.
See e.g. \cite{StromingerLectures,StromingerIRrevisited,Gomez2,Sever,Porrati} for recent discussions and reviews. The soft photon clouds render the LGT charges of Faddeev-Kulish (FK) states independent of the momenta of the bare charged particles \cite{Sever}. The LGT charges depend only on the net electric charge of the bare particles (and the angle dependent constants at infinity), and so the conservation laws can be trivially satisfied \cite{StromingerIRrevisited,Sever}.

Thus any scattering process in QED inevitably leads to a final state with an infinite number of soft photons. Our goal is to study the entanglement of the hard particles with the soft photons produced in a typical process such as electron-electron scattering, and calculate the entanglement entropy perturbatively. Even though we focus on a particular process, we expect the main conclusions to be applicable to other (perturbative) scattering processes in quantum electrodynamics as well as gravity.

Depending on the sensitivity of the detector, we impose an energy cutoff $E < m_e$, where $m_e$ is the electron mass, in terms of which we decompose the {\it incoming} and {\it outgoing} Hilbert spaces into hard and soft factors
\begin{equation}
{\cal{H}}= {\cal{H}}_H \times {\cal{H}}_S
\end{equation}
where ${\cal{H}}_H$ comprises hard electron, positron and photon states with energy greater than $E$, and ${\cal{H}}_S$ of soft photon states with total energy less than $E$. The initial state is taken to be a two-electron dressed FK state, but we also discuss and compare with perturbative Fock basis computations. The final state will be an entangled state in ${\cal{H}}_H \times {\cal{H}}_S$, as determined by the $S$-matrix. By restricting the incoming energy, we may exclude the possibility of having more than two charged particles in the final state. In addition, we shall distinguish between photons produced as a result of radiation and photons present in the clouds accompanying the outgoing charged particles.

Note that apart from the infrared reference scale $E$ used to decompose the Hilbert space into soft and hard factors, we also have the following infrared energy scales: i)$\lambda$ is the infrared cutoff scale, eventually to be taken to zero. Any logarithmic IR divergences in physical quantities will be displayed as powers of $\log\lambda$. ii) $E_d$ characterizes the energy of the soft photons present in the clouds accompanying the incoming and outgoing charged particles. We set $\lambda < E_d < E$, taking $E_d$ to be sufficiently small so that the leading soft photon theorems can be applied to simplify various dressed amplitudes (see below). iii) $\Lambda$ (which can be taken to be of order $E_d$) is an infrared scale characterizing soft virtual photons. Eventually we take the limit $\lambda \to 0$, keeping the ratios $E_d/E$ and $E/\Lambda$ fixed. We would like to investigate the behavior of the entanglement entropy as the reference scale $E$ approaches the lower infrared scales $E_d$ and $\lambda$, in perturbation theory, as well as the $\lambda \to 0$ limit.

In the rest of this section we review some properties of the FK construction, which will be useful for the entanglement entropy computations among the soft and hard particles. Throughout we work in the Lorenz gauge. For notation and conventions, see \ref{A1}. 

%%%%%%%%%%%%%%%%%%%%%%%%%%%%%%%%%%%%%%%%%%%%%%%%%%%%%%%%%%%%%%%%%%%%%%%%
\subsection{Faddeev-Kulish states}
%%%%%%%%%%%%%%%%%%%%%%%%%%%%%%%%%%%%%%%%%%%%%%%%%%%%%%%%%%%%%%%%%%%%%%%%%%
\noindent
The FK dressing is effected via the action of $e^{R_f}$,
where
\begin{equation}
R_f=\int~\frac{d^3\vec{p}}{(2\pi)^3}~\hat{\rho}(\vec{p})~\int_\lambda^{E_d}\frac{d^3\vec{k}}{(2\pi)^3}~\frac{1}{(2\omega_{\vec{k}})^{1/2}}~\left(f(\vec{k},\,\vec{p})\cdot a^{\dagger}(\vec{k}) - h.c.\right) \label{FKdressing}
\end{equation}
and
\begin{equation}
\hat{\rho}(\vec{p})=\sum_s b^{s\, \dagger}(\vec{p})b^s(\vec{p}) - d^{s\, \dagger}(\vec{p})d^s(\vec{p})
\end{equation}
is the charge density operator; $b^{s\, \dagger}(\vec{p})$ and $d^{s\, \dagger}(\vec{p})$ are electron and positron creation operators, respectively -- $\vec{p}$ is the momentum and $s$ the spin polarization; $a_r^{\dagger}(\vec{k})$ create photons with momentum $\vec{k}$ and polarization vector $\epsilon_r^{\mu}(\vec{k})$, $r=0,\dots,3$,
and 
\begin{equation}
f(\vec{k},\,\vec{p})\cdot a^{\dagger}(\vec{k})=\sum_r f^{\mu}(\vec{k},\,\vec{p})\epsilon^*_{r\mu}(\vec{k})a^{\dagger}_r(\vec{k}) \label{f1}
\end{equation}
with
\begin{equation}
f^{\mu}(\vec{k},\,\vec{p})= e\left(\frac{p^{\mu}}{pk} - c^{\mu}\right)e^{-ipk\, t_0/p^0},\,\,\, c^{\mu}=\left(-\frac{1}{2k^0},\,\frac{\vec{k}}{2(k^0)^2}\right) \label{f2}
\end{equation}
The FK operator is unitary. Notice that the dressing function $f^{\mu}(\vec{k},\,\vec{p})$ is singular as the photon momentum $\vec{k}$ vanishes. We will carry all computations keeping the infrared cutoff scale $\lambda$ finite, taking the $\lambda \to 0$ limit at the end. Here also, $t_0$ is a time reference scale and $c^{\mu}$ is a null vector, $c^2=0$, satisfying $ck=1$. Because of the latter property, the function $f^{\mu}(\vec{k},\,\vec{p})$ is transverse, $fk=0$. So only allowable admixtures of timelike and longitudinal photons are present, in accordance with the Lorenz gauge condition. In particular, these do not contribute to the $S$-matrix elements (as well as to expectation values of gauge invariant quantities)\footnote{Dressed states satisfy the Gupta-Bleuler condition: $\left[a_0(\vec{k})-a_3(\vec{k})\right]\ket{\Psi}=0$.}, and thus we may also restrict the sum in \ref{f1} to transversely polarized photons. The limits of integration in \ref{FKdressing} insure that only soft photons, with energies below the infrared reference scale $E_d$, are present in the cloud. As the integrand is dominated by low momenta, taking $E_d < 1/t_0$, we may approximate the phase $e^{-ipk\, t_0/p^0}$ in \ref{f2} with unity.

%%%%%%%%%%%%%%%%%%%%%%%%%%%%%%%%%%%%%%%%%%%%%%%%%%%%%%%%%%%%%%%%%%%%%%%%%%%%%%%%%
\subsection{Dressed electron}
%%%%%%%%%%%%%%%%%%%%%%%%%%%%%%%%%%%%%%%%%%%%%%%%%%%%%%%%%%%%%%%%%%%%%
\noindent
For example, consider a bare single electron particle state
\begin{equation}
|\vec{p},\, s \rangle= \sqrt{2 E_{\vec{p}}}\, b^{s\, \dagger}(\vec{p})|0\rangle 
\end{equation}
%(with momentum $\vec{p}$ and spin polarization $s$).
The corresponding dressed state takes a product form
%in ${\cal{H}}_H \times {\cal{H}}_S$
\begin{equation}
|\vec{p},\,s\rangle_{\rm dressed}= |\vec{p},\,s\rangle\times e^{\int_\lambda^{E_d}\frac{d^3\vec{k}}{(2\pi)^3}~\frac{1}{(2\omega_{\vec{k}})^{1/2}}~\left(f(\vec{k},\,\vec{p})\cdot a^{\dagger}(\vec{k}) - h.c.\right)}|0\rangle\label{coherent1}
\end{equation}
Thus the charged particle is accompanied by a photon cloud described by a normalized coherent state. For finite nonzero $\lambda$, the coherent state can also be  written in the following useful form
\begin{equation}
|f_{\vec{p}}\rangle = {\cal{N}}_{\vec{p}}~\,e^{\int_\lambda^{E_d}\frac{d^3\vec{k}}{(2\pi)^3}~\frac{1}{(2\omega_{\vec{k}})^{1/2}}~f(\vec{k},\,\vec{p})\cdot a^{\dagger}(\vec{k})}|0\rangle \label{coherent2}
\end{equation}
The normalization factor ${\cal{N}}_{\vec{p}}$ is given by  
\begin{equation}
{\cal{N}}_{\vec{p}}=e^{-\frac{1}{2} \int_\lambda^{E_d}\frac{d^3\vec{q}}{(2\pi)^3}~\frac{1}{2\omega_{\vec{q}}}~f^{\mu}(\vec{q},\,\vec{p})f^{*}_{\mu}(\vec{q},\,\vec{p}) } \label{normalization}
\end{equation}
The exponent can easily be computed
\begin{equation}
\frac{1}{2} \int_\lambda^{E_d}\frac{d^3\vec{q}}{(2\pi)^3}~\frac{1}{2\omega_{\vec{q}}}~f^{\mu}(\vec{q},\,\vec{p})f^{*}_{\mu}(\vec{q},\,\vec{p}) = \frac{e^2}{8\pi^2}~\ln{\left(\frac{E_d}{\lambda}\right)}~I(v)
\end{equation}
%$$
%\frac{1}{2} \int_\lambda^{E_d}\frac{d^3\vec{q}}{(2\pi)^3}~\frac{e^2}{2\omega_{\vec{q}}}~\left(\frac{p^2}{(pq)^2}-2~\frac{cp}{pq}\right)
%= \frac{e^2}{4} \ln{\left(\frac{E_d}{\lambda}\right)}~\int_{-1}^1 \frac{dx}{(2\pi)^2}\left(\frac{-m_e^2}{\left(p^0 - |\vec{p}|\,x\right)^2}+~\frac{p^0 +|\v%ec{p}|\,x }{p^0 - |\vec{p}|\,x}\right)
%$$
where $v= |\vec{p}|/p^0$ is the velocity of the electron and
\begin{equation}
I(v) =-2~+~v^{-1}\ln \left(\frac{1+ v}{1-v}\right) \label{kinematical1}
\end{equation}
is a non-negative kinematical factor. In particular, for small $v$, $I(v)=2v^2/3+\dots$\, . As $v \to 1$, $I(v)$ grows logarithmically. Setting
\begin{equation}
{\cal{A}}_{\vec{p}} = \frac{e^2}{8\pi^2}~I(v)
\end{equation}
we get
\begin{equation}
{\cal{N}}_{\vec{p}}=\left(\frac{\lambda}{E_d}\right)^{{\cal{A}}_{\vec{p}}}
\end{equation}
and so ${\cal{N}}_{\vec{p}}$ vanishes in the limit $\lambda \to 0$ (in which case \ref{coherent2} cannot be used).

Let us compute the number of photons in this state. Using standard coherent state algebra, this is given by
\begin{equation}
\langle f_{\vec{p}}| N_{ph} |f_{\vec{p}}\rangle = \int_{\lambda}^{E_d} \frac{d^3 \vec{q}}{(2\pi)^3}~\frac{1}{2\omega_{\vec{q}}}~ f^{\mu}(\vec{q},\, \vec{p})f^*_{\mu}(\vec{q},\, \vec{p})=\frac{e^2}{4\pi^2} \ln{\left(\frac{E_d}{\lambda}\right)}~I(v)
\end{equation}
Thus the cloud contains an infinite number of soft photons in the limit $\lambda \to 0$. On the other hand, the energy of the state is given by
\begin{equation}
\langle f_{\vec{p}}| H_{ph} |f_{\vec{p}}\rangle = \frac{1}{2}~ \int_{\lambda}^{E_d} \frac{d^3 \vec{q}}{(2\pi)^3}~ f^{\mu}(\vec{q},\, \vec{p})f^*_{\mu}(\vec{q},\, \vec{p})=\frac{e^2}{4\pi^2}~I(v)~(E_d-\lambda)
\end{equation}
For generic values of the electron velocity, this is a small fraction of the infrared scale $E_d$. Therefore, the coherent cloud of photons is in the soft part of the Hilbert space ${\cal{H}}_S$.

The mean value of the cloud momentum is also interesting. It is given by
\begin{equation}
\langle f_{\vec{p}}| \vec{P}_{ph} |f_{\vec{p}}\rangle =  \int_{\lambda}^{E_d} \frac{d^3 \vec{q}}{(2\pi)^3}~\frac{\vec{q}}{2\omega_{\vec{q}}}~ f^{\mu}(\vec{q},\, \vec{p})f^*_{\mu}(\vec{q},\, \vec{p})=\frac{e^2}{8\pi^2}~(E_d-\lambda)~\left[\frac{3}{v}~I(v)-v~I(v)-2~v\right]~\hat{p}
\end{equation}
As the electron velocity approaches the speed of light, the energy and the magnitude of the cloud momentum grow logarithmically and become equal. Notice that both the energy and the momentum remain appreciably much smaller than the energy and the momentum of the electron. As $v\to 0$, they become vanishingly small, albeit the momentum approaches zero faster. 

Finally let us compute the electromagnetic field associated with the cloud.
%\footnote{In addition there is the Lienard Wiechert field produced by the back-reacting electron.}
The expectation value of the gauge potential in the coherent state is
\begin{equation}
\bar{A}_\mu(x)=\langle f_{\vec{p}}| A_{\mu}(x) |f_{\vec{p}}\rangle =  \int_{\lambda}^{E_d} \frac{d^3 \vec{q}}{(2\pi)^3}~\frac{1}{2\omega_{\vec{q}}}~\left(f_{\mu}(\vec{q},\, \vec{p})~e^{iqx}~+~f^*_{\mu}(\vec{q},\, \vec{p})~e^{-iqx}\right)
\end{equation}
As noted in \cite{Sotaro}, the terms in $\bar{A}_\mu(t_0,\,\vec{x})$ that are independent of the the null vector $c^{\mu}$ reproduce asymptotically the Lienard Wiechert gauge potential associated with the moving electron. 
%Since the dressing function $f_{\mu}(\vec{q},\, \vec{p})$ is transverse, $\bar{A}_\mu(x)$ satisfies the Lorenz condition, $\partial%_\mu\bar{A}^\mu(x)=0$, and the equation $\Box\bar{A}_\mu(x)=0$.
The electric field is
\begin{equation}
\vec{E}=  \int_{\lambda}^{E_d} \frac{d^3 \vec{q}}{(2\pi)^3}~\frac{ie}{2\omega_{\vec{q}}}~\left(\frac{\hat{q}-\vec{v}}{1-\hat{q}\cdot\vec{v}}~+~\vec{v}~-~(1+\hat{q}\cdot\vec{v})~\frac{\hat{q}}{2}\right)~e^{iq(x-pt_0/p^0)}~+~h.c.
\end{equation}
The last term in the parentheses is the contribution of the null vector $c^{\mu}$.
%Using the asymptotic expression for plane waves
%\begin{equation}
%\lim_{r \to \infty,\,u\,\rm{fixed}} e^{iqx} = \frac{2\pi i}{|\vec{q}|~r}\left[e^{-i(|\vec{q}|-i\epsilon)(u+2r)}\delta^2(\hat{x}+\ha%t{q})-e^{-i(|\vec{q}|-i\epsilon)u}\delta^2(\hat{x}-\hat{q})\right]
%\end{equation}
%where $u=t-r$, we can obtain the asymptotic behavior of the electric field near future null infinity (${\cal{I}}^+$):
%\begin{equation}
%\lim_{r \to \infty,\,u\,\rm{fixed}}\vec{E}=\frac{e}{4\pi^2 \,r}~\left(\frac{\hat{x}-\vec{v}}{1-\hat{x}\cdot\vec{v}}-\hat{x}\right)~%\frac{\sin\left[E_d(t_0~-~\hat{x}\cdot\vec{v}~t_o~-~u)\right]}{t_0(1-\hat{x}\cdot\vec{v})-u}
%\end{equation}
%This is a low frequency transverse, $\vec{E}\cdot \hat{x}=0$, radiative field. Note also that when $\hat{x}=\hat{v}$, the asymptoti%c electric field vanishes.

For the case of multielectron/positron states, $\alpha=\{e_i,\, \vec{p}_i,\, s_i\}$, the resulting coherent state $|f_\alpha\rangle$ can be obtained if we replace the function $f^{\mu}(\vec{k},\,\vec{p})$ in expressions
\ref{coherent1} and \ref{coherent2} with 
\begin{equation}
f_\alpha^{\mu}(\vec{k})=\sum_{i \in \alpha}~e_i~\left(\frac{p_i^{\mu}}{p_ik} - c^{\mu}\right)~e^{-ip_ik\, t_0/p_i^0} 
\end{equation}
where $e_i$ is the charge and $p_i$ is the momentum of the $i$th particle. When $t_0=0$, the second term is equal to $Q_\alpha c^{\mu}$, where $Q_\alpha$ is the total charge of $\alpha$. In particular, the terms proportional to $c^{\mu}$ vanish for states with zero net charge. For simplicity, we choose to set the phases $e^{-ip_ik\, t_0/p_i^0}$ to unity for the following calculations \cite{Gomez2,Chung}.    

In \ref{A2} we compute the normalization factor ${\cal{N}}_\alpha$ for the photon coherent state associated with the state $\alpha$. We also compute the overlap between coherent photon states, corresponding to generic charged states $\alpha=\{e_i,\, \vec{p}_i,\,s_i\}$ and $\beta=\{e_i^{\prime}, \,\vec{p}_i^{\,\prime},\,s_i^{\,\prime}\}$. Let us call the $\beta$ particles outgoing and the $\alpha$ particles incoming, and define $\eta_i$ to be $+1$ for all outgoing particles and $-1$ for all incoming particles. Then for the cases of interest $Q_\alpha=Q_\beta$ and to all orders in the electron charge, we find 
\begin{equation}
\langle f_\beta|f_\alpha\rangle = \left(\frac{\lambda}{E_d}\right)^{{\cal{B}}_{\beta\alpha}}\label{braketf}
\end{equation}
where
\begin{equation}
{\cal{B}}_{\beta\alpha}=-\frac{1}{16\pi^2}~\sum_{ij}~\eta_i\,\eta_j\,e_i\,e_j~v_{ij}^{-1}~ \ln \left(\frac{1+ v_{ij}}{1-v_{ij}}\right)\label{B}
\end{equation}
and 
\begin{equation}
v_{ij}=\left[1-\frac{m_i^2\,m_j^2}{(p_i\cdot p_j)^2}\right]^{1/2}
\end{equation}
is (the magnitude of) the relative velocity of particle $j$ with respect to $i$.
The sums are over all outgoing and incoming particles. When the momenta of the multicharged particle states $\beta$ and $\alpha$ differ, ${\cal{B}}_{\beta\alpha}$ is nonzero and positive \cite{Weinberg}. Then to all orders in the electron charge, the overlap $\langle f_\beta|f_\alpha\rangle$ vanishes in the $\lambda \to 0$ limit.

%%%%%%%%%%%%%%%%%%%%%%%%%%%%%%%%%%%%%%%%%%%%%%%%%%%%%%%%%%%
\subsection{The Faddeev-Kulish $S$-matrix}
%%%%%%%%%%%%%%%%%%%%%%%%%%%%%%%%%%%%%%%%%%%%%%%%%%%%%%%%%
\noindent
Next consider a scattering process $\alpha \to \beta$. We first consider cases for which there are no soft photons with energy less than $E_d$ in the initial and final states (beyond the ones specified by the dressing operator). We compute the $S$-matrix element between the incoming/outgoing dressed states, following \cite{Chung}:
\begin{equation}
{\tilde S}_{\beta\alpha}=_{d}\langle\beta|S|\alpha\rangle_d
\end{equation}
We also write
\begin{equation}
S_{\beta\alpha}=\langle\beta|S|\alpha\rangle
\end{equation}
for the $S$-matrix element between the corresponding undressed states. Expanding the exponential operators of the coherent photon states, we obtain
\begin{equation}
   {\tilde S}_{\beta\alpha}=
   {\cal{N}}_\beta~{\cal{N}}_\alpha
   \sum_{m,\,n\,=0}^\infty~\frac{1}{m!\,n!}~\langle\beta|~\prod_{l=1}^m~\int_{\lambda}^{E_d} \frac{d^3\vec{q}_l}{(2\pi)^3}~\frac{f^*_\beta(\vec{q}_l)\cdot a(\vec{q}_l)}{(2\omega_{\vec{q}_l})^{1/2}}~S~\prod_{s=1}^n~\int_{\lambda}^{E_d} \frac{d^3\vec{k}_s}{(2\pi)^3}~\frac{f_\alpha(\vec{k}_s)\cdot a^{\dagger}(\vec{k}_s)}{(2\omega_{\vec{k}_s})^{1/2}}~|\alpha\rangle \label{Sdressed}
   \end{equation}
   
Each term in \ref{Sdressed} is given in terms of scattering amplitudes with $n$ incoming soft photons and $m$ outgoing soft photons. These amplitudes are weighted by $1/m!\,n!$. It is always possible that a number $l$, $0\leq l \leq \rm{min}(m,\,n)$, of these soft photons do not interact with the electrons. Then $n^\prime=n-l$ soft photons are absorbed by an external electron line, and $m^{\prime}=m-l$ are emitted by an external electron line.

The $l$ noninteracting soft photons contribute a factor given by
\begin{equation}
l! ~ \left(\int_\lambda^{E_d}\frac{d^3\vec{q}}{(2\pi)^3}~\frac{1}{2\omega_{\vec{q}}}~f_\alpha^{\mu}(\vec{q})f^{*}_{\beta\,\mu}(\vec{q})\right)^l
\end{equation}
Notice that the sum over photon polarizations -- we restrict the sum over transversely polarized photons ($r=1,2$) -- yields
\begin{equation}
\sum_r \epsilon_{r\mu}(\vec{q})\epsilon^*_{r\nu}(\vec{q})=\eta_{\mu\nu}-q_{\mu}c_{\nu}-q_{\nu}c_{\mu}
\end{equation}   
and we have used the fact that the dressing functions are transverse $f^*_{\beta}q=f_{\alpha}q=0$.
Letting the energy scale $E_d$ to be sufficiently small, we can 
obtain the contributions of the $n^\prime$ and $m^\prime$ interacting soft photons by using the following leading soft theorems \cite{Bloch,Low1,GellMann,Low2,Yennie,BK,Weinberg}
\begin{equation}
\lim_{|\vec{q}|\to 0}~(2\omega_{\vec{q}})^{1/2}~\langle\beta|a_r(\vec{q})~S~|\alpha\rangle=\left(\sum_{i\in\beta}~ \frac{e_i\,p_i \cdot \epsilon_r^*(\vec{q})}{p_i \cdot q}~-~\sum_{i\in \alpha}~ \frac{e_i\, p_i \cdot \epsilon_r^*(\vec{q})}{p_i \cdot q}\right)~\langle\beta|~S~|\alpha\rangle
\end{equation}
and (by CPT invariance)
\begin{equation}
\lim_{|\vec{k}|\to 0}~(2\omega_{\vec{k}})^{1/2}~\langle\beta|~S~a^{\dagger}_r(\vec{k})~|\alpha\rangle=-\left(\sum_{i\in\beta}~ \frac{e_i\,p_i \cdot \epsilon_r(\vec{k})}{p_i \cdot k}~-~\sum_{i\in \alpha}~ \frac{e_i\, p_i \cdot \epsilon_r(\vec{k})}{p_i \cdot k}\right)~\langle\beta|~S~|\alpha\rangle
\end{equation}

Then $\tilde{S}_{\beta\alpha}$ can be expressed as a sum over all possible $(l, m^{\prime}, n^{\prime})$ configurations, after taking into account all weight factors, including the fact that there are $(n^{\prime}+l)!/n^\prime!\,l!$ ways to choose $l$ photons from the initial $n$ soft photons, and likewise $(m^{\prime}+l)!/m^\prime!\,l!$ ways to choose $l$ photons from the final $m$ soft photons. In all we get
$$
{\tilde S}_{\beta\alpha}=
{{\cal{N}}_\beta~{\cal{N}}_\alpha}
~\sum_{l,\,m^{\prime},\,n^{\prime}\,=\,0}^\infty ~\frac{1}{(m^\prime +l)!\,(n^{\prime}+l)!}~\frac{(m^{\prime}+l)!}{m^\prime!\,l!}~\frac{(n^{\prime}+l)!}{n^\prime!\,l!}~l! ~ \left(\int_\lambda^{E_d}\frac{d^3\vec{q}}{(2\pi)^3}~\frac{1}{2\omega_{\vec{q}}}~f_\alpha^{\mu}(\vec{q})f^{*}_{\beta\,\mu}(\vec{q})\right)^l
$$
$$
\times~\left[\int_{\lambda}^{E_d} \frac{d^3\vec{q}}{(2\pi)^32\omega_{\vec{q}}}~\sum_{i\in{\{\beta,\,\alpha\}}}~\eta_i\,e_i~\left(\frac{f^*_\beta(\vec{q})\cdot p_i}{p_i\cdot q}-f^*_\beta(\vec{q})\cdot c\right)\right]^{m^{\prime}}
$$
\begin{equation}
\times~\left[-\int_{\lambda}^{E_d} \frac{d^3\vec{k}}{(2\pi)^32\omega_{\vec{k}}}~\sum_{i\in{\{\beta,\,\alpha\}}}~\eta_i\,e_i~\left(\frac{f_\alpha(\vec{k})\cdot p_i}{p_i\cdot k}-f_\alpha(\vec{k})\cdot c\right)\right]^{n^{\prime}}~S_{\beta\alpha}
\end{equation}
The last two lines are the contributions of the $n^\prime$ and $m^\prime$ interacting soft photons. It is easy to see that the terms proportional to $c$ vanish by charge conservation ($Q_\alpha=Q_\beta$). After canceling combinatorial factors, it is easy to see that all three series exponentiate to give
\begin{equation}
   {\tilde S}_{\beta\alpha}=~
   {{\cal{N}}_\beta~{\cal{N}}_\alpha}~e^{\int_\lambda^{E_d}\frac{d^3\vec{q}}{(2\pi)^3}~\frac{1}{2\omega_{\vec{q}}}~f_\alpha^{\mu}(\vec{q})f^{*}_{\beta\,\mu}(\vec{q})}~
   ~e^{\int_{\lambda}^{E_d} \frac{d^3\vec{k}}{(2\pi)^32\omega_{\vec{k}}}\sum_{ij}~\eta_i\,\eta_j\,e_i\,e_j~\frac{p_i~p_j}{(p_ik)~(p_jk)}}~S_{\beta\alpha}
\end{equation}

The first three factors combine to produce $\langle f_\beta|f_\alpha\rangle$ given by \ref{braketf}. In the second exponential, the $ij$ sums are over all outgoing and incoming particles. The exponent is given by 
\begin{equation}
\int_{\lambda}^{E_d} \frac{d^3\vec{k}}{(2\pi)^32\omega_{\vec{k}}}\sum_{ij}~\eta_i\,\eta_j\,e_i\,e_j~\frac{p_j~p_i}{(p_jk)~(p_ik)}=\ln\left(\frac{E_d}{\lambda}\right)~(2{\cal{B}}_{\beta\alpha})
\end{equation}
where ${\cal{B}}_{\beta\alpha}$ is the positive kinematical factor given by \ref{B}. Therefore
\begin{equation}
   {\tilde S}_{\beta\alpha}=
   ~\langle f_\beta|f_\alpha\rangle~
   ~\left(\frac{E_d}{\lambda}\right)^{2{\cal{B}}_{\beta\alpha}}~S_{\beta\alpha}~=~\left(\frac{E_d}{\lambda}\right)^{{\cal{B}}_{\beta\alpha}}~S_{\beta\alpha}
\end{equation}
On the other hand, as shown in e.g. \cite{Weinberg}, exponentiation of virtual infrared divergences gives
\begin{equation}
S_{\beta\alpha}=\left(\frac{\lambda}{\Lambda}\right)^{{\cal{B}}_{\beta\alpha}}~e^{i\phi_{\beta\alpha}}~S^{(\Lambda)}_{\beta\alpha}
\end{equation}
with $S^{(\Lambda)}_{\beta\alpha}$ the usual $S$-matrix amplitude without virtual soft photons with momentum below the infrared scale $\Lambda$. The phase $\phi_{\alpha\beta}$ is real \cite{Weinberg} and does not contribute to the square of the amplitudes or the corresponding rates. So
\begin{equation}
{\tilde S}_{\beta\alpha}=~\left(\frac{E_d}{\Lambda}\right)^{{\cal{B}}_{\beta\alpha}}~e^{i\phi_{\beta\alpha}}~S^{(\Lambda)}_{\beta\alpha}
\end{equation}
is finite (generically nonzero and free of infrared divergences). In the limit $\lambda \to 0$, we keep the ratio $E_d/\Lambda$ finite. (We may also choose to set $\Lambda=E_d$).

\subsubsection{Single soft photon production}
Now let us add a single soft photon $\gamma$, of momentum $\vec{q}_\gamma$ and polarization vector $\epsilon_{r\mu}(\vec{q}_\gamma)$ ($|\vec{q}_\gamma|<E_d$), in the final state:
\begin{equation}
{\tilde S}_{\beta\gamma,\,\alpha}=_{d}\langle\beta\gamma|S|\alpha\rangle_d
\end{equation}
The case $|\vec{q}_\gamma|>E_d$ is covered by the previous analysis. Such amplitudes in QED and gravity were recently studied in \cite{Choi1,Choi2}.

To calculate the $S$-matrix element, we first note that
\begin{equation}
|\beta\gamma\rangle_d = \left(|\beta\gamma\rangle~-~f_\beta^{*\,\mu}(\vec{q}_\gamma)\epsilon_{r\mu}(\vec{q}_\gamma)~|\beta\rangle\right)\times|f_\beta\rangle
\end{equation}
as obtained by acting with the FK operator $e^{R_f}$ on the undressed state $|\beta\gamma\rangle=a_r^\dagger(\vec{q}_\gamma)\ket{\beta}$. Notice that the trivial part of the $S$-matrix element, given by the overlap $\,_{d}\langle\beta\gamma|\alpha\rangle_d$, vanishes:
\begin{equation}
\,_{d}\langle\beta\gamma|\alpha\rangle_d=\left(f_\alpha(\vec{q}_\gamma)-f_\beta(\vec{q}_\gamma)\right)\cdot\epsilon_{r}^*(\vec{q}_\gamma)~\langle f_\beta|f_\alpha\rangle~\langle\beta|\alpha\rangle=\left(f_\alpha(\vec{q}_\gamma)-f_\alpha(\vec{q}_\gamma)\right)\cdot\epsilon_{r}^*(\vec{q}_\gamma)=0
\end{equation}
(since $\langle\beta|\alpha\rangle=\delta_{\beta\alpha}$). So two states which differ by an extra photon with energy less than $E_d$ (beyond the ones in the dressing) are orthogonal, and so distinguishable. As a result, only the nontrivial part of the $S$-matrix contributes to this matrix element.

In all, ${\tilde S}_{\beta\gamma,\,\alpha}$ can be written as a sum of two parts
\begin{equation}
{\tilde S}_{\beta\gamma,\,\alpha}~=~{\tilde S}_{\beta\gamma,\,\alpha}^{(1)}~+~{\tilde S}_{\beta\gamma,\,\alpha}^{(2)}
\end{equation}
where
\begin{equation}
{\tilde S}_{\beta\gamma,\,\alpha}^{(1)}~=~-f_\beta(\vec{q}_\gamma)\cdot\epsilon_{r}^*(\vec{q}_\gamma)~\tilde{S}_{\beta\alpha}~=~-f_\beta(\vec{q}_\gamma)\cdot\epsilon_{r}^*(\vec{q}_\gamma)~\left(\frac{E_d}{\lambda}\right)^{{\cal{B}}_{\beta\alpha}}~S_{\beta\alpha}
\end{equation}
and
$$
{\tilde S}_{\beta\gamma,\,\alpha}^{(2)}~=~{\cal{N}}_\beta~{\cal{N}}_\alpha~\sum_{m,\,n\,=0}^\infty~\frac{(2\omega_\gamma)^{1/2}}{m!\,n!}
$$
\begin{equation}
\times~\langle\beta|~a_r(\vec{q}_\gamma)~\prod_{l=1}^m~\int_{\lambda}^{E_d} \frac{d^3\vec{q}_l}{(2\pi)^3}~\frac{f^*_\beta(\vec{q}_l)\cdot a(\vec{q}_l)}{(2\omega_{\vec{q}_l})^{1/2}}~S~\prod_{s=1}^n~\int_{\lambda}^{E_d} \frac{d^3\vec{k}_s}{(2\pi)^3}~\frac{f_\alpha(\vec{k}_s)\cdot a^{\dagger}(\vec{k}_s)}{(2\omega_{\vec{k}_s})^{1/2}}~|\alpha\rangle 
\end{equation}
For the second part, we note that there are two contributions, depending on whether the extra outgoing soft photon (annihilated by $a_r(\vec{q}_\gamma)$) is interacting. {\it Feynman diagrams in which this extra soft photon is joined to an external electron line yield a net contribution}
\begin{equation}
S^{(2)}_1=\left(\frac{E_d}{\lambda}\right)^{{\cal{B}}_{\beta\alpha}}~S_{\beta\gamma,\,\alpha}=\left(\frac{E_d}{\lambda}\right)^{{\cal{B}}_{\beta\alpha}}~S_{\beta\alpha}~\left(\sum_{i\in\beta}~ \frac{e_i\,p_i \cdot \epsilon_r^*(\vec{q}_\gamma)}{p_i \cdot q_\gamma}~-~\sum_{i\in \alpha}~ \frac{e_i\, p_i \cdot \epsilon_r^*(\vec{q}_\gamma)}{p_i \cdot q_\gamma}\right)~+~\dots
\end{equation}
We have used the soft theorem for sufficiently small $|\vec{q}_\gamma|$. The ellipses stand for smooth, nonsingular terms in the limits $\lambda,\,|\vec{q}_\gamma|\to 0$ \footnote{The subleading ${\cal{O}}(\omega_\gamma^0)$ terms obey a universal relation \cite{Low1,GellMann,Low2,BK,Duca,LPS}. At the one loop level, corrections that are logarithmic in the photon frequency can arise \cite{HHW,Bern,BHHW,Sahoo}. These corrections do not affect the leading perturbative computation of the entanglement entropy in \ref{s4}. Notice also that such a logarithmic singularity in the amplitude would be integrable. In particular, it leads to suppressed contributions, of the order $E_d\log E_d$, in various physical quantities, where we integrate over the soft photon momentum.}. Since $Q_\alpha = Q_\beta$, this gives
\begin{equation}
\tilde{S}^{(2)}_1~=~\left(\frac{E_d}{\lambda}\right)^{{\cal{B}}_{\beta\alpha}}~S_{\beta\alpha}~\left(f_\beta(\vec{q}_\gamma) \cdot \epsilon_r^*(\vec{q}_\gamma)~-~f_\alpha(\vec{q}_\gamma) \cdot \epsilon_r^*(\vec{q}_\gamma)\right)~+~\dots
\end{equation}

{\it Let us now consider the case for which the extra soft photon is not interacting}. Let the total number of outgoing noninteracting soft photons be $1+l$, and likewise for the incoming ones. Then $n^\prime=n-l-1$ soft photons are absorbed by an external electron line, and $m^{\prime}=m-l$ are emitted by an external electron line. Now the noninteracting soft photons contribute a factor given by
\begin{equation}
(1+l)~l! ~f_\alpha(\vec{q}_\gamma) \cdot \epsilon_r^*(\vec{q}_\gamma)~ \left(\int_\lambda^{E_d}\frac{d^3\vec{q}}{(2\pi)^3}~\frac{1}{2\omega_{\vec{q}}}~f_\alpha^{\mu}(\vec{q})f^{*}_{\beta\,\mu}(\vec{q})\right)^l
\end{equation}
For the interacting soft photons we must apply the soft theorems as before.

We then sum over all possible $(l, m^{\prime}, n^{\prime})$ configurations, after taking into account all weight factors. Notice that there are $(n^{\prime}+l+1)!/n^\prime!\,(l+1)!$ ways to choose $l+1$ photons from the initial $n$ soft photons, and likewise $(m^{\prime}+l)!/m^\prime!\,l!$ ways to choose $l$ photons from the final $m$ soft photons. In all we get
$$
\tilde{S}^{(2)}_2=
{{\cal{N}}_\beta~{\cal{N}}_\alpha}
~\sum_{l,\,m^{\prime},\,n^{\prime}\,=\,0}^\infty ~\frac{1}{(m^\prime +l)!\,(n^{\prime}+l+1)!}~\frac{(m^{\prime}+l)!}{m^\prime!\,l!}~\frac{(n^{\prime}+l+1)!}{n^\prime!\,(l+1)!}
$$
$$
\times~(1+l)~l! ~f_\alpha(\vec{q}_\gamma) \cdot \epsilon_r^*(\vec{q}_\gamma) ~ \left(\int_\lambda^{E_d}\frac{d^3\vec{q}}{(2\pi)^3}~\frac{1}{2\omega_{\vec{q}}}~f_\alpha^{\mu}(\vec{q})f^{*}_{\beta\,\mu}(\vec{q})\right)^l
$$
$$
\times~\left[\int_{\lambda}^{E_d} \frac{d^3\vec{q}}{(2\pi)^32\omega_{\vec{q}}}~\sum_{i\in{\{\beta,\,\alpha\}}}~\eta_i\,e_i~\left(\frac{f^*_\beta(\vec{q})\cdot p_i}{p_i\cdot q}-f^*_\beta(\vec{q})\cdot c\right)\right]^{m^{\prime}}
$$
\begin{equation}
\times~\left[-\int_{\lambda}^{E_d} \frac{d^3\vec{k}}{(2\pi)^32\omega_{\vec{k}}}~\sum_{i\in{\{\beta,\,\alpha\}}}~\eta_i\,e_i~\left(\frac{f_\alpha(\vec{k})\cdot p_i}{p_i\cdot k}-f_\alpha(\vec{k})\cdot c\right)\right]^{n^{\prime}}~S_{\beta\alpha}
\end{equation}
%The last two lines are the contributions of the the $n^\prime$ and $m^\prime$ interacting soft photons. It is easy to see that the terms proportional to $c$ vanish by charge conservation ($Q_\alpha=Q_\beta$).
After canceling combinatorial factors as before, it is easy to see that all three series exponentiate to give
\begin{equation}
   \tilde{S}^{(2)}_2~=~\left(\frac{E_d}{\lambda}\right)^{{\cal{B}}_{\beta\alpha}}~S_{\beta\alpha}~f_\alpha(\vec{q}_\gamma) \cdot \epsilon_r^*(\vec{q}_\gamma)
   \end{equation}
and so
\begin{equation}   
   {\tilde S}_{\beta\gamma,\,\alpha}^{(2)}~=~ \tilde{S}^{(2)}_1 ~+~\tilde{S}^{(2)}_2~=~\left(\frac{E_d}{\lambda}\right)^{{\cal{B}}_{\beta\alpha}}~S_{\beta\alpha}~f_\beta(\vec{q}_\gamma) \cdot \epsilon_r^*(\vec{q}_\gamma)~+~\dots
\end{equation}

Therefore, adding the two parts together, we find that all singular terms, in the limits $\lambda,\,|\vec{q}_\gamma| \to 0$, cancel:
\begin{equation}
{\tilde S}_{\beta\gamma,\,\alpha}~=~F_{\beta\alpha}(\vec{q}_\gamma,\,\epsilon_r(\vec{q}_\gamma))
\end{equation}
Here $F_{\beta\alpha}(\vec{q}_\gamma,\,\epsilon_r(\vec{q}_\gamma))$ is a smooth function as $\lambda,\,|\vec{q}_\gamma| \to 0$. In fact, it has been shown that by appropriately correcting the dressing function to subleading order in the soft photon momentum (and to leading order in the electron charge), this function is of order $E_d$ \cite{Choi2}. So the dressing suppresses the emission of soft photons with energy $\omega_\gamma < E_d$, at least at tree level. We conclude that the dressed amplitude ${\tilde S}_{\beta\gamma,\,\alpha}$ is nonsingular, and suppressed when $\omega_\gamma < E_d$. {\it This motivates us to distinguish between low frequency photons with frequencies in the range $E_d<\omega_\gamma<E$, comprising the soft part of the emitted radiation, and soft photons present in the clouds accompanying the outgoing charged particles}. It would be interesting to see if the suppression of ${\tilde S}_{\beta\gamma,\,\alpha}$ persists at the one loop level \cite{Choi2}, since then corrections logarithmic in the soft photon frequency appear. One would need to consider $e^2$ corrections to the dressing function for this task.

%%%%%%%%%%%%%%%%%%%%%%%%%%%%%%%%%%%%%%%%%%%%%%%%%%%%%%%%%%%%%%%%%%
\section{Discretization}\label{s3}
%%%%%%%%%%%%%%%%%%%%%%%%%%%%%%%%%%%%%%%%%%%%%%%%%%%%%%%%%%%%%%%%%%%%
\noindent
For the entanglement entropy computation, we replace infinite space with a large box of size $L$ (volume $V=L^3$) and impose periodic boundary conditions for the fields. The momenta are quantized as
\begin{equation}
\vec{k} = \frac{2\pi}{L}(n_1,\,n_2,\,n_3) 
\end{equation}
We also rescale the annihilation/creation operators
\begin{equation}
a_r(\vec{k}) \to V^{1/2}~\tilde{a}_r(\vec{k})
\end{equation}
so that for the discrete system, the commutation relations read
\begin{equation}
[\tilde{a}_r(\vec{k}),\,\tilde{a}_{r^\prime}^\dagger(\vec{k^\prime})]=\delta_{rr^\prime}\delta_{\vec{k}\,\vec{k}^\prime}
\end{equation}
Here $\delta_{rr^\prime}$ and $\delta_{\vec{k}\,\vec{k}^\prime}$ are Kronecker deltas. We restrict to transversely polarized photons. The single particle states
\begin{equation}
\tilde{a}_{r^\prime}^\dagger(\vec{k})|0\rangle
\end{equation}
are unit normalized. The IR cutoff scale $\lambda$ is naturally taken to be equal to $2\pi/L$. We will drop the tildes for simplicity.

%Note that apart from the infrared cutoff scale $\lambda$, we also have the following infrared energy scales: i) $E$ is the reference scale used to %decompose the Hilbert space into soft and hard parts, and it is set by the sensitivity of the detector. ii) $E_d$ characterizes the energy of the s%oft photons present in the clouds accompanying the initial charged particles. Recall that the FK prescription is not unique. We set $\lambda < E_d %< E$, but take $E_d$ to be sufficiently small, so that the soft photon theorems can be applied to simplify various amplitudes (see the previous sec%tion). iii) $\Lambda$ is an infrared scale characterizing soft virtual photons. Eventually we take the continuum limit $\lambda \to 0$, keeping the% ratios $E/\Lambda$, $E_d/\Lambda$ and $E_d/E$ fixed. One may also set the scales $E$ and $E_d$ equal, but generically they can be different.

Consider now an initially undressed two electron state $|\beta\rangle=|e_me_n\rangle$. The indices stand for both momentum and polarization.
The effect of dressing yields
\begin{equation}
|\beta\rangle_d=|e_me_n\rangle_H \times |f_\beta\rangle_S
\end{equation}
where $|f_\beta\rangle_S$ is the
coherent state describing the cloud of soft photons. For the discrete system, this is given by
\begin{equation}
|f_{\beta}\rangle_S = U_{\beta}|0\rangle_S={\cal{N}}_{\beta}~e^{A_\beta^\dagger}~|0\rangle_S
\end{equation}
with
\begin{equation}
U_{\beta} = e^{(A_\beta^\dagger - A_\beta)},\,\,\,\,\, A_\beta=\sum_{\omega_{\vec{k}}<{E_d}}~\frac{1}{(2\,V\,\omega_{\vec{k}})^{1/2}}~f^*_{\beta}(\vec{k})\cdot a(\vec{k}) 
\end{equation}
and
\begin{equation}
{\cal{N}}_{\beta}=e^{-\frac{1}{2} \sum_{\omega_{\vec{k}}<{E_d}}~\frac{1}{2\,V\,\omega_{\vec{k}}}~f_{\beta}^{\mu}(\vec{k})f^{*}_{\beta\,\mu}(\vec{k}) } 
\end{equation}
As shown before, $|f_\beta\rangle$ is in the soft part of the Hilbert space ${\cal{H}}_S$.

Next we form the ket-bra operator
\begin{equation}
|\beta\rangle_{d}\langle\beta^\prime|_{d}
\end{equation}
(with $|\beta^\prime\rangle$ a different two electron state). Tracing over the soft part of the Hilbert space gives
\begin{equation}
\Tr_{{\cal{H}}_S}\left(|\beta\rangle_{d}\langle\beta^\prime|_{d}\right)=|\beta\rangle_{H}\langle\beta^\prime|_{H}~\langle f_{\beta^\prime}|f_\beta\rangle\label{trace1}
\end{equation}
The overlap $\langle f_{\beta^\prime}|f_\beta\rangle$ has been computed in \ref{braketf}, in the continuum limit. In particular, when $\beta\ne\beta^\prime$ and to all orders in the electron charge, the overlap vanishes in the strict $\lambda \to 0$ limit. For any superposition of dressed states, tracing over the soft part of the Hilbert space leads to decoherence and an almost diagonal density matrix \cite{Carney3}.

Now suppose that we add a single soft photon to the undressed state $|\beta\rangle$. Let the soft photon momentum be $\vec{q}$ ($|\vec{q}|<E$) and denote the polarization vector by $\epsilon_{r\mu}(\vec{q})$:
\begin{equation}
|\beta\gamma(r,\,\vec{q})\rangle=a^\dagger_r(\vec{q})|\beta\rangle
\end{equation}
Decomposing the corresponding dressed state in ${\cal{H}}_H \times {\cal{H}}_S$ yields
\begin{equation}
|\beta\gamma\rangle_d=|\beta\rangle_H \times \left(U_{\beta}~a^\dagger_r(\vec{q})~|0\rangle_S\right)=|\beta\rangle_H \times
%\left({\cal{N}}_{\beta}~e^{A_\beta^{\dagger}}~\large(a^\dagger_r(\vec{q})-[A_\beta,\,a^\dagger_r(\vec{q})]\large)~|0\rangle_S\right)
\left(a^\dagger_r(\vec{q})-[A_\beta,\,a^\dagger_r(\vec{q})]\right)~|f_\beta\rangle_S
\end{equation}
Using this expression, we can readily calculate the partial traces:
\begin{equation}
\Tr_{{\cal{H}}_S}\left(|\beta\gamma\rangle_{d}\langle\beta^\prime|_{d}\right)=|\beta\rangle_{H}\langle\beta^\prime|_{H}~\langle f_{\beta^\prime}|f_\beta\rangle~\left([A_{\beta^\prime},\,a^\dagger_r(\vec{q})] - [A_\beta,\,a^\dagger_r(\vec{q})]\right)
\end{equation}
If $E_d<|\vec{q}|<E$, the commutators vanish. On the other hand, if $|\vec{q}|<E_d$, the commutators are nontrivial and give
\begin{equation}
\Tr_{{\cal{H}}_S}\left(|\beta\gamma\rangle_{d}\langle\beta^\prime|_{d}\right)=|\beta\rangle_{H}\langle\beta^\prime|_{H}~\langle f_{\beta^\prime}|f_\beta\rangle~\frac{1}{(2V\omega_{\vec{q}})^{1/2}}~\left(f_{\beta^\prime}^*(\vec{q})-f_\beta^*(\vec{q})\right)\cdot \epsilon_r(\vec{q}) \label{trace2}
\end{equation}
Notice that this vanishes for the diagonal cases $\beta=\beta^\prime$. Also, the function $f_\beta^{\mu}(\vec{q})$ is of order $e$. Next we compute
$$
\Tr_{{\cal{H}}_S}\left(|\beta\gamma\rangle_{d}\langle\beta^\prime\gamma^\prime|_{d}\right)=|\beta\rangle_{H}\langle\beta^\prime|_{H}~\langle f_{\beta^\prime}|f_\beta\rangle
$$
\begin{equation}
\times ~ \left\{\delta_{rr^\prime}\delta_{\vec{q}\vec{q}^\prime}+\left([a_{r^\prime}(\vec{q}^\prime),\, A_\beta^\dagger]-[a_{r^\prime}(\vec{q}^\prime),\, A_{\beta^\prime}^\dagger]\right)~\left([A_{\beta^\prime},\,a^\dagger_r(\vec{q})] - [A_\beta,\,a^\dagger_r(\vec{q})]\right)\right\}
\end{equation}
If both $|\vec{q}|,\,|\vec{q}^\prime|<E_d$, we obtain
$$
\Tr_{{\cal{H}}_S}\left(|\beta\gamma\rangle_{d}\langle\beta^\prime\gamma^\prime|_{d}\right)~=~|\beta\rangle_{H}\langle \beta^\prime|_{H}~\langle f_{\beta^\prime}|f_\beta\rangle
$$
\begin{equation}
\times~\left\{\delta_{rr^\prime}\delta_{\vec{q}\vec{q}^\prime}+\frac{1}{(2V\omega_{\vec{q}^\prime})^{1/2}(2V\omega_{\vec{q}})^{1/2}}~\left(f_\beta(\vec{q}^\prime)-f_{\beta^\prime}(\vec{q}^\prime)\right)\cdot \epsilon_{r^\prime}(\vec{q}^\prime)~\left(f_{\beta^\prime}(\vec{q})-f_\beta(\vec{q})\right)\cdot \epsilon_{r}(\vec{q})\right\} \label{trace3}
\end{equation}
In a similar way, we can compute partial traces for the cases in which two or more soft photons are present in the initially undressed states. There will be contributions that are higher order in the function $f_\beta^{\mu}(\vec{q})$.

\section{Scattering with dressed states and entanglement entropy}\label{s4}
\noindent
The incoming state is taken to be
\begin{equation}
|\psi\rangle_{in}=|e_ie_j\rangle_d=|e_ie_j\rangle_H \times |f_\alpha\rangle_S
\end{equation}
We will also adopt the notation $|\alpha\rangle=|e_ie_j\rangle$.
Notice that this is a product state and so there is no entanglement between the soft and hard degrees of freedom \footnote{Had we started with a superposition of dressed states, there would be entanglement between the soft and hard degrees of freedom \cite{Carney4}.}. Entanglement occurs as a result
of scattering. In particular the initial density matrix, including tracing over the undetectable soft photon clouds, is pure:
\begin{equation}
\Tr_{{\cal{H}}_S}\left(|\psi\rangle_{in}\langle\psi|_{in}\right)=|\alpha\rangle_H\langle\alpha|_H
\end{equation}

The out state is given in terms of the $S$-matrix by
\begin{equation}
|\psi\rangle_{out}=S|\psi\rangle_{in}=(1+iT)|\alpha\rangle_d
\end{equation}
For simplicity, we restrict the incoming energy so that electron/positron pair production is forbidden, and so only two charged particles are present in the final state. Since the $S$-matrix is unitary, we have
\begin{equation}
i(T-T^{\dagger})=-T^{\dagger}T \label{unitarity}
\end{equation}

Inserting a complete basis of dressed states, $\ket{\psi}_{out}$ can be written as 
\begin{equation}
|\psi\rangle_{out}=|\alpha\rangle_d~+~\tilde{A}_{\beta\alpha}|\beta\rangle_d~+~\tilde{B}_{\beta\gamma,\,\alpha}|\beta\gamma\rangle_d+\dots
\end{equation}
where $\tilde{A}_{\beta\alpha}=_d\bra{\beta}iT\ket{\alpha}_d$ and $\tilde{B}_{\beta\gamma,\,\alpha}=_d\bra{\beta\gamma}iT\ket{\alpha}_d$ are $S$-matrix elements between dressed states. Summation over the final state electron and photon indices $\beta$ and $\gamma$, respectively, is implied. The leading contributions in $\tilde{A}_{\beta\alpha}$ are of order $e^2$ and in $\tilde{B}_{\beta\gamma,\,\alpha}$ of order $e^3$. The ellipses stand for higher order contributions, arising from states with two or more photons. The associated density matrix is (no sum over $\alpha$)
$$
\ket{\psi}_{out}\bra{\psi}_{out}=~\ket{\alpha}_d\bra{\alpha}_d~+~ \left(\tilde{A}_{\beta\alpha}\ket{\beta}_d~+~\tilde{B}_{\beta\gamma,\,\alpha}\ket{\beta\gamma}_d~+~\dots\right)\bra{\alpha}_d
$$
$$
+~\ket{\alpha}_d\left(\tilde{A}^*_{\beta^\prime\alpha}\bra{\beta^\prime}_d~+~\tilde{B}^*_{\beta^\prime\gamma^\prime,\,\alpha}\bra{\beta^\prime\gamma^\prime}_d~+~\dots\right)
$$
$$
+~\tilde{A}_{\beta\alpha}\tilde{A}^*_{\beta^\prime\alpha}\ket{\beta}_d\bra{\beta^\prime}_d~+~\tilde{B}_{\beta\gamma,\,\alpha}\tilde{B}^*_{\beta^\prime\gamma^\prime,\,\alpha}\ket{\beta\gamma}_d\bra{\beta^\prime\gamma^\prime}_d
$$
\begin{equation}
~+~\tilde{B}_{\beta\gamma,\,\alpha}\tilde{A}^*_{\beta^\prime\alpha}\ket{\beta\gamma}_d\bra{\beta^\prime}_d~+~\tilde{A}_{\beta\alpha}\tilde{B}^*_{\beta^\prime\gamma^\prime,\,\alpha}\ket{\beta}_d\bra{\beta^\prime\gamma^\prime}_d ~+~\dots
\end{equation}

As we already remarked, we will discuss two partial traces and the associated density matrices: 1) over all soft photons in ${\cal{H}}_S$ and 2) over soft photons with frequencies in the range $E_d<\omega_\gamma<E$, comprising the soft part of the emitted radiation. The latter is motivated by the fact that the amplitude for the emission of soft photons with energy less than $E_d$ is suppressed \cite{Choi2} -- see the discussion at the end of \ref{s2}. In the first case, the reduced density matrix is an operator in ${\cal{H}}_H$. Since the second case does not prescribe tracing over soft cloud photons, we obtain an operator acting on the space of physical asymptotic states.  

\subsection{Tracing over all soft photons}
First we trace over all soft photons, including those in the clouds. The reduced density matrix
\begin{equation}
\rho_H=\Tr_{{\cal{H}}_S}\left(\ket{\psi}_{out}\bra{\psi}_{out}\right)
\end{equation}
can be readily obtained using expressions \ref{trace1}, \ref{trace2} and \ref{trace3}. It takes the following form
$$
\rho_H=\ket{\alpha}_H\bra{\alpha}_H + \left(C_{\beta}\ket{\beta}_H~+~\sum_{\omega_\gamma>E}C_{\beta\gamma}\ket{\beta\gamma}_H+\dots\right)\bra{\alpha}_H
$$
$$
+\ket{\alpha}_H\left(C^*_{\beta^\prime}\bra{\beta^\prime}_H~+~\sum_{\omega_{\gamma^\prime}>E}C^*_{\beta^\prime\gamma^\prime}\bra{\beta^\prime\gamma^\prime}_H+\dots\right)
$$
$$
+D_{\beta,\,\beta^\prime}\ket{\beta}_H\bra{\beta^\prime}_H~+~\sum_{\omega_\gamma,\omega_{\gamma^\prime}>E}D_{\beta\gamma,\,\beta^\prime\gamma^\prime} \ket{\beta\gamma}_H\bra{\beta^\prime\gamma^\prime}_H 
$$
\begin{equation}
+ \sum_{\omega_\gamma>E}D_{\beta\gamma,\,\beta^\prime}\ket{\beta\gamma}_H\bra{\beta^\prime}_H~+~\sum_{\omega_{\gamma^\prime}>E}D^*_{\beta^\prime\gamma^\prime,\,\beta}\ket{\beta}_H\bra{\beta^\prime\gamma^\prime}_H  ~+~ \dots
\end{equation}
where 
\begin{equation}
C_{\beta} = \bra{f_\alpha}\ket{f_\beta}~\left(\tilde{A}_{\beta\alpha}~+~\sum_{\omega_\gamma<{E_d}}~\frac{1}{(2V\omega_\gamma)^{1/2}}~\tilde{B}_{\beta\gamma,\,\alpha}\left(f_\alpha^*(\vec{q}_\gamma)-f_\beta^*(\vec{q}_\gamma)\right)\cdot\epsilon(\gamma)~+~\dots\right)\label{C1}
\end{equation}
$$
$$
\begin{equation}
C_{\beta\gamma}=\bra{f_\alpha}\ket{f_\beta}~\tilde{B}_{\beta\gamma,\,\alpha}~+~\dots \label{C2}
\end{equation}
$$
$$
$$
D_{\beta,\,\beta^{\prime}}/\bra{f_{\beta^\prime}}\ket{f_\beta}=\tilde{A}_{\beta\alpha}\tilde{A}^*_{\beta^\prime\alpha} ~+~\sum_{\omega_\gamma<{E_d}}~\frac{1}{(2V\omega_\gamma)^{1/2}}~\tilde{B}_{\beta\gamma,\,\alpha}\tilde{A}^*_{\beta^\prime\alpha}\left(f_{\beta^\prime}^*(\vec{q}_\gamma)-f_\beta^*(\vec{q}_\gamma)\right)\cdot\epsilon(\gamma)
$$
\begin{equation}
%+~\bra{f_{\beta^\prime}}\ket{f_\beta}~
+~\sum_{\omega_{\gamma^\prime}<{E_d}}\frac{1}{(2V\omega_{\gamma^\prime})^{1/2}}~\tilde{B}^*_{\beta^\prime\gamma^\prime,\,\alpha}\tilde{A}_{\beta\alpha}\left(f_{\beta}(\vec{q}_{\gamma^\prime})-f_{\beta^\prime}(\vec{q}_{\gamma^\prime})\right)\cdot\epsilon^*(\gamma^\prime)+\sum_{\omega_{\gamma}<E}\tilde{B}_{\beta\gamma,\,\alpha}\tilde{B}^*_{\beta^\prime\gamma,\,\alpha}~+~\dots \label{C3}
\end{equation}
$$
$$
\begin{equation}
D_{\beta\gamma,\,\beta^\prime\gamma^\prime}=\bra{f_{\beta^\prime}}\ket{f_\beta}~\tilde{B}_{\beta\gamma,\,\alpha}\tilde{B}^*_{\beta^\prime\gamma^\prime,\,\alpha}~+~\dots\label{C4}
\end{equation}
$$
$$
\begin{equation}
D_{\beta\gamma,\,\beta^\prime}=\bra{f_{\beta^\prime}}\ket{f_\beta}~\tilde{B}_{\beta\gamma,\,\alpha}\tilde{A}^*_{\beta^\prime\alpha}~+~\dots\label{C5}
\end{equation}

The matrix elements of $\rho_H$ are given in terms of dressed amplitudes, which are free of any IR divergences (at least perturbatively),
as well as overlaps of coherent photon states describing the soft clouds. The diagonal elements are proportional to inclusive Bloch-Nordsieck type rates associated with dressed box states, and they are free of any IR divergences in $\lambda$, order by order in perturbation theory \footnote{As we discuss in \ref{continuum}, they scale inversely with (powers of) the volume in the continuum (large volume) limit.}.
For example
\begin{equation}
D_{\beta,\,\beta}=\tilde{A}_{\beta\alpha}\tilde{A}^*_{\beta\alpha}~+~\sum_{\omega_{\gamma}<E}\tilde{B}_{\beta\gamma,\,\alpha}\tilde{B}^*_{\beta\gamma,\,\alpha}~+~\dots \label{Ddiagonal} 
\end{equation}  
is proportional to the rate for the transition of the initial dressed state $\ket{\alpha}_d$ to $\ket{\beta}_d$, and any number of photons with total energy less than $E$.

The off diagonal elements, e.g. $D_{\beta,\,\beta^\prime}$ ($\beta\ne\beta^\prime$), are proportional to the overlap $\bra{f_{\beta^\prime}}\ket{f_\beta}$, which, at any finite order in perturbation theory, induces logarithmic divergences in $\lambda$ (via its perturbative expansion at finite $\lambda$ -- see \ref{bracketf2} below). Thus generically, at any finite order in perturbation theory, the off diagonal elements are nonzero and must be taken into account. {\it To all orders in the electron charge}, these IR logarithmic terms exponentiate. As a result, when the momenta of the two-electron particle states $\beta$ and $\beta^\prime$ differ, the overlap $\bra{f_{\beta^\prime}}\ket{f_\beta}$ to all orders vanishes in the strict $\lambda \to 0$ limit. {\it Therefore, to all orders in perturbation theory, the density matrix assumes an almost diagonal form in the continuum limit, exhibiting decoherence} \cite{Carney3}. In the following, we keep the volume of the box and the infrared cutoff $\lambda$ finite in order to regularize the entanglement entropy, {\it working at finite order in perturbation theory and taking into account the contributions of the off diagonal elements.}
%of $\rho_H$ for small but finite $\lambda$
We would like to investigate if in the continuum limit, the entanglement entropy is free of any IR logarithmic divergences in $\lambda$, order by order in perturbation theory.

This behavior of the off diagonal elements is reminiscent of the behavior of the conventional Fock basis amplitudes (with a finite number of photons in the initial and final states). At any finite order in perturbation theory, these amplitudes are nonzero, containing logarithmic divergences in $\lambda$ due to virtual soft photons. Their contributions must be taken into account in the perturbative calculation of the inclusive cross sections. Notice, however, that to all orders in perturbation theory, the virtual infrared divergences exponentiate, causing the individual Fock basis amplitudes to vanish. IR logarithmic divergences in $\lambda$ appear also due to real soft photon emission. The inclusive cross sections are free of any IR divergences, order by order in perturbation theory. The IR divergences due to virtual soft photons and real soft photon emission cancel against each other in this case \cite{Weinberg}. 

We can extract the analogous Fock basis computation, where the initial state is taken to be a state of two bare electrons, by setting the function $f_\beta^{\mu}(\vec{q})$ to be zero and replacing the dressed amplitudes with conventional Fock basis amplitudes. At any finite order in perturbation theory, the off diagonal elements are nonzero and contain logarithmic divergences in $\lambda$. {\it To all orders in perturbation theory however,} the IR divergences exponentiate, leading to the vanishing of the off diagonal elements of the corresponding density matrix in the continuum, $\lambda \to 0$ limit \cite{Carney2}. The diagonal elements are given in terms of Bloch-Nordsieck rates, and so they are free of IR divergences order by order in perturbation theory. As in the dressed case, we fix $\lambda$ and work at finite order in perturbation theory, taking into account the contributions of the off diagonal elements to the entanglement entropy. The continuum $\lambda \to 0$ limit is taken at the end. 

Some more comments are in order:
\begin{itemize}
\item
For the Fock basis case, it is clear that the entanglement between the soft and hard parts of the Hilbert space arises due to soft photon emission. The entanglement entropy is of order $e^6$ \footnote{More precisely, the leading entanglement entropy is of order $e^6\ln{e^6}$. The entanglement entropy is not analytic in $e$ at $e=0$.}, with Feynman diagrams involving the emission of a single soft photon contributing at leading order. Likewise for the dressed case, the entanglement entropy is of order $e^6$. At lower orders, the density matrix assumes a product form, and so it is pure. 
\item
The leading contributions in $C_{\beta}$ are of order $e^2$, in $C_{\beta\gamma}$ of order $e^3$, in $D_{\beta,\,\beta^\prime}$ of order $e^4$, in $D_{\beta\gamma,\,\beta^\prime}$ of order $e^5$ and in $D_{\beta\gamma,\,\beta^\prime\gamma^\prime}$ of order $e^6$.
\item
The last three terms in  $D_{\beta,\,\beta^\prime}$, see equation \ref{C3}, are of order $e^6$. The ellipses include terms of higher order than $e^6$, which do not contribute to the entanglement entropy at leading order. Likewise, the ellipses in $D_{\beta\gamma,\,\beta^\prime\gamma^\prime}$, see \ref{C4}, include terms of higher order than $e^6$, which can be ignored at leading order.    
\item
The second term in $C_{\beta}$ (equation \ref{C1}) is of order $e^4$ and vanishes when $\beta=\alpha$. Contributions from two or more photon states are proportional to the matrix elements $\tilde{B}_{\beta{\gamma_1}{\gamma_2},\dots,\gamma_i\dots,\,\alpha}$ and products of the differences $f_{\alpha}^\mu(\vec{q}_i)-f_{\beta}^\mu(\vec{q}_i)$, and so they also vanish when $\beta=\alpha$. Only the first term contributes in $C_{\alpha}$: $C_{\alpha}=\tilde{A}_{\alpha\alpha}$ to all orders.
\item
As we will see, $C_{\alpha}+C_{\alpha}^*$ is of order $e^4$ by unitarity. This result considerably simplifies the leading order computation of the entanglement entropy.
%\item
%The Fock basis computation can be recovered by setting the dressing functions $f_{\alpha}^\mu(\vec{q}_i)$ and $f_{\beta}^\mu(\vec{q}_j)$ to zero.   
\end{itemize}

\subsection{Perturbative analysis to order $e^6$}
We proceed to compute the Renyi entropies
\begin{equation}
S_{m}=\frac{1}{1-m}\log \Tr (\rho_H)^m
\end{equation}
for integer $m \ge 2$, to leading order in perturbation theory ($e^6$). Had the density matrix $\rho_H$ corresponded to a pure state (all eigenvalues zero but one eigenvalue equal to one), the Renyi entropies would vanish. So they measure the degree of entanglement and the information carried by the soft photons. The entanglement entropy 
\begin{equation}
S_{ent}=-\Tr \rho_H \log \rho_H
\end{equation}
can be written as an infinite series of the Renyi entropies for integer $m$ \footnote{We can also obtain the entanglement entropy in the limit $m \to 1$: $\lim_{m \to 1} S_{m} = S_{ent}$.}:
\begin{equation}
S_{ent}=\sum_{n=1}^{\infty}\sum_{m=0}^{n}\frac{(n-1)!}{(n-m)!~m!}~(-1)^m~e^{-mS_{m+1}} \label{entR}
\end{equation}

Let us set
\begin{equation}
\rho_H = \rho_0 + {\varepsilon}
\end{equation}
where $\rho_0 = \ket{\alpha}_H\bra{\alpha}_H$. 
Then since $\Tr \rho_H=\Tr \rho_0=1$, $\Tr \varepsilon =0$. Indeed computing the trace explicitly, we obtain 
$$
\Tr \varepsilon = C_\alpha + C_\alpha^* + \sum_{\beta}\left(D_{\beta,\,\beta} + \sum_{\omega_\gamma>E} D_{\beta\gamma,\, \beta\gamma}\right) + \dots
$$
\begin{equation}
=\tilde{A}_{\alpha\alpha}+\tilde{A}_{\alpha\alpha}^* + \sum_{\beta}\tilde{A}_{\beta\alpha}\tilde{A}_{\beta\alpha}^* + \sum_{\beta\gamma}\tilde{B}_{\beta\gamma,\,\alpha}\tilde{B}_{\beta\gamma,\,\alpha}^* + \dots=_d\bra{\alpha}i(T-T^\dagger)+T^\dagger T\ket{\alpha}_d=0 \label{unitarity2}
\end{equation}
The last equation follows from unitarity (see \ref{unitarity}).

Now $\varepsilon$ is of order $e^2$. So to obtain the leading contribution to the trace of $\rho_H^m$ ($m>2$) and to the corresponding Renyi entropy (which is of order $e^6$), we need to expand $\rho_H^m$ to the cubic order in $\varepsilon$. The fact that $\rho_0^2=\rho_0$ and the cyclic property of the trace limit the number of structures we need to consider.

At the linear level, we need only compute $\varepsilon \rho_0$ and its trace:
$$
\varepsilon \rho_0~=~C_{\beta}\ket{\beta}_H\bra{\alpha}_H ~+~ \sum_{\omega_\gamma>E}C_{\beta\gamma}\ket{\beta\gamma}_H\bra{\alpha}_H ~+~ C_\alpha^*\ket{\alpha}_H\bra{\alpha}_H
$$
\begin{equation}
+~D_{\beta,\,\alpha}\ket{\beta}_H\bra{\alpha}_H~+~\sum_{\omega_\gamma>E}D_{\beta\gamma,\,\alpha}\ket{\beta\gamma}_H\bra{\alpha}_H~+~\dots
\end{equation}
$$
$$
\begin{equation}
\Tr(\varepsilon \rho_0)~=~C_{\alpha}~+~C_{\alpha}^*~+~D_{\alpha,\,\alpha}~=~\tilde{A}_{\alpha\alpha}~+~\tilde{A}_{\alpha\alpha}^* ~+~\tilde{A}_{\alpha\alpha}\tilde{A}_{\alpha\alpha}^*~+~\left(\sum_{\omega_{\gamma}<E}\tilde{B}_{\alpha\gamma,\,\alpha}\tilde{B}^*_{\alpha\gamma,\,\alpha}\right)~+~\dots 
\end{equation}
The ellipses in the trace include terms of order higher than $e^6$ and can be dropped to leading order in the entanglement entropy.

At the quadratic level, it suffices to consider the following structures: $\varepsilon^2$, $\varepsilon^2\rho_0$ and $\varepsilon\rho_0\varepsilon\rho_0$. First we get:
$$
\varepsilon^2 ~=~ \left(C_\beta C_\beta^*~+~\sum_{\omega_\gamma>E}C_{\beta\gamma}C_{\beta\gamma}^* \right)~\ket{\alpha}_H\bra{\alpha}_H 
$$
$$
+~\left(C_\beta C_\alpha~+~D_{\beta,\, \beta^\prime}C_{\beta^\prime}\right) ~\ket{\beta}_H\bra{\alpha}_H ~+~ \left(C_\beta^* C_\alpha^*~+~C_{\beta^\prime}^*D_{\beta^\prime,\,\beta}\right) ~\ket{\alpha}_H\bra{\beta}_H 
$$
$$
+~\sum_{\omega_\gamma>E}C_{\beta\gamma}C_{\alpha}~\ket{\beta\gamma}_H\bra{\alpha}_H~+~C_{\beta\gamma}^*C_{\alpha}^*~\ket{\alpha}_H\bra{\beta\gamma}_H
$$
$$
+~\left(C_\beta C_{\beta^\prime}^*~+~D_{\beta,\,\alpha}C_{\beta^\prime}^*~+~D_{\alpha,\,\beta^\prime}C_{\beta} \right)~\ket{\beta}_H\bra{\beta^\prime}_H ~+~ \sum_{\omega_\gamma, \omega_{\gamma^\prime}>E}C_{\beta\gamma}C_{\beta^\prime\gamma^\prime}^*~\ket{\beta\gamma}_H\bra{\beta^\prime\gamma^\prime}_H
$$
\begin{equation}
+~\sum_{\omega_\gamma>E}C_{\beta\gamma}C_{\beta^\prime}^*~\ket{\beta\gamma}_H\bra{\beta^\prime}_H~+~C_{\beta\gamma}^*C_{\beta^\prime}~\ket{\beta^\prime}_H\bra{\beta\gamma}_H~+~\dots
\end{equation}
Taking the trace yields
$$
\Tr \varepsilon^2 = C_\alpha^2 + C_\alpha^{*\, 2} + \sum_\beta \left(2D_{\alpha,\, \beta}C_\beta +
2D_{\beta,\,\alpha}C_{\beta}^*+ 2C_\beta C_\beta^*~+~2\sum_{\omega_\gamma>E}C_{\beta\gamma}C_{\beta\gamma}^*\right) + \dots
$$
$$
= \tilde{A}_{\alpha\alpha}^2 + \tilde{A}_{\alpha\alpha}^{*\, 2}+2\sum_{\beta} |\bra{f_\beta}\ket{f_\alpha}|^2\left(1+\tilde{A}_{\alpha\alpha}+\tilde{A}_{\alpha\alpha}^*\right)\tilde{A}_{\beta\alpha}\tilde{A}_{\beta\alpha}^*
$$
$$
+2\sum_{\beta}\sum_{\omega_\gamma<{E_d}}\frac{|\bra{f_\beta}\ket{f_\alpha}|^2}{(2V\omega_\gamma)^{1/2}} \left[ \tilde{A}_{\beta\alpha}\tilde{B}_{\beta\gamma,\,\alpha}^*\left(f_\alpha(\vec{q}_\gamma)-f_\beta(\vec{q}_\gamma)\right)\cdot \epsilon^*(\gamma)+\tilde{A}_{\beta\alpha}^*\tilde{B}_{\beta\gamma,\,\alpha}\left(f_\alpha^*(\vec{q}_\gamma)-f_\beta^*(\vec{q}_\gamma)\right)\cdot \epsilon(\gamma)\right]
$$
\begin{equation}
+2\sum_\beta\sum_{\omega_\gamma>E}|\bra{f_\beta}\ket{f_\alpha}|^2\tilde{B}_{\beta\gamma,\,\alpha}\tilde{B}_{\beta\gamma,\,\alpha}^* +\dots
\end{equation}
In the last line, the ellipses stand for terms of higher order than $e^6$. The second line vanishes when the dressing function is set to zero, and so it is absent in the Fock basis computation.

Next we calculate $\varepsilon^2 \rho_0$:
$$
\varepsilon^2\rho_0 ~=~ \left(C_{\alpha}^{*\, 2}~+~C_\beta C_\beta^*~+~\sum_{\omega_\gamma>E}C_{\beta\gamma}C_{\beta\gamma}^* ~+~C_{\beta}^*D_{\beta,\,\alpha}\right)~\ket{\alpha}_H\bra{\alpha}_H 
$$
$$
+~\left(C_\beta C_\alpha~+~D_{\beta,\, \beta^\prime}C_{\beta^\prime}~+~C_\beta C_{\alpha}^*~+~D_{\beta,\,\alpha}C_{\alpha}^*~+~D_{\alpha,\,\alpha}C_{\beta}\right) ~\ket{\beta}_H\bra{\alpha}_H 
$$
\begin{equation}
+~\sum_{\omega_\gamma>E}\left(C_{\beta\gamma}C_{\alpha}~+~C_{\beta\gamma}C_{\alpha}^*\right)~\ket{\beta\gamma}_H\bra{\alpha}_H~+~\dots
\end{equation}
For the trace we obtain
$$
\Tr (\varepsilon^2\rho_0) = C_\alpha^2 + C_\alpha^{*\, 2} + C_{\alpha}C_{\alpha}^*+ D_{\alpha,\,\alpha} (C_\alpha + C_{\alpha}^*)
$$
$$
+ \sum_\beta \left(D_{\alpha,\, \beta}C_\beta +
D_{\beta,\,\alpha}C_{\beta}^*+ C_\beta C_\beta^*~+~\sum_{\omega_\gamma>E}C_{\beta\gamma}C_{\beta\gamma}^*\right) + \dots
$$
$$
= \tilde{A}_{\alpha\alpha}^2 + \tilde{A}_{\alpha\alpha}^{*\, 2}+\left(1+\tilde{A}_{\alpha\alpha}+\tilde{A}_{\alpha\alpha}^*\right) \tilde{A}_{\alpha\alpha}\tilde{A}_{\alpha\alpha}^*+\sum_{\beta} |\bra{f_\beta}\ket{f_\alpha}|^2\left(1+\tilde{A}_{\alpha\alpha}+\tilde{A}_{\alpha\alpha}^*\right)\tilde{A}_{\beta\alpha}\tilde{A}_{\beta\alpha}^*
$$
$$
+\sum_{\beta}\sum_{\omega_\gamma<{E_d}}\frac{|\bra{f_\beta}\ket{f_\alpha}|^2}{(2V\omega_\gamma)^{1/2}} \left[ \tilde{A}_{\beta\alpha}\tilde{B}_{\beta\gamma,\,\alpha}^*\left(f_\alpha(\vec{q}_\gamma)-f_\beta(\vec{q}_\gamma)\right)\cdot \epsilon^*(\gamma)+\tilde{A}_{\beta\alpha}^*\tilde{B}_{\beta\gamma,\,\alpha}\left(f_\alpha^*(\vec{q}_\gamma)-f_\beta^*(\vec{q}_\gamma)\right)\cdot \epsilon(\gamma)\right]
$$
\begin{equation}
+\sum_\beta\sum_{\omega_\gamma>E}|\bra{f_\beta}\ket{f_\alpha}|^2\tilde{B}_{\beta\gamma,\,\alpha}\tilde{B}_{\beta\gamma,\,\alpha}^* +\dots
\end{equation}

Notice the appearance of off diagonal elements in the perturbative expansion for both $\Tr \varepsilon^2$ and $\Tr \varepsilon^2\rho_0$. As we remarked before, at any finite order in perturbation theory, the off diagonal elements are nonzero and contain IR logarithmic divergences in $\lambda$. Both of these traces contribute to the Renyi and the entanglement entropies to leading order ($e^6$), and we would like to investigate whether the logarithmic divergences cancel.

Finally for $\varepsilon\rho_0\varepsilon\rho_0$ and its trace we get:
$$
\varepsilon \rho_0\varepsilon\rho_0~=~\left[(C_{\alpha}+C_{\alpha}^*+D_{\alpha,\,\alpha})C_{\beta}+(C_{\alpha}+C_{\alpha}^*)D_{\beta,\,\alpha}\right]\ket{\beta}_H\bra{\alpha}_H 
$$
\begin{equation}
~+~ \sum_{\omega_\gamma>E}(C_{\alpha}+C_{\alpha}^*)C_{\beta\gamma}\ket{\beta\gamma}_H\bra{\alpha}_H~+~C_\alpha^*(C_\alpha + C_\alpha^* + D_{\alpha,\,\alpha})\ket{\alpha}_H\bra{\alpha}_H ~+~\dots
\end{equation}
$$
$$
$$
\Tr(\varepsilon \rho_0\varepsilon\rho_0)=(C_\alpha + C_\alpha^*)(C_\alpha + C_\alpha^*+2D_{\alpha,\,\alpha})+\dots
$$
\begin{equation}
=(\tilde{A}_{\alpha\alpha}+\tilde{A}_{\alpha\alpha}^*)(\tilde{A}_{\alpha\alpha}+\tilde{A}_{\alpha\alpha}^* +2\tilde{A}_{\alpha\alpha}\tilde{A}_{\alpha\alpha}^*)+\dots
\end{equation}
This trace vanishes to order $e^6$ by unitarity.

To cubic order, it is sufficient to compute $\varepsilon^3$, $\varepsilon^3\rho_0$, $\varepsilon^2\rho_0\varepsilon\rho_0$ and
$\varepsilon\rho_0\varepsilon\rho_0\varepsilon\rho_0$.
As we will show, these traces are vanishing to order $e^6$, and so they do not contribute to the entanglement entropy at leading order. Indeed, we find
$$
\varepsilon^3 ~=~ (C_\alpha + C_\alpha^*) C_\beta C_\beta^* ~\ket{\alpha}_H\bra{\alpha}_H~+~C_{\beta^\prime} (C_\beta C_\beta^* + C_\alpha^2) \ket{\beta^\prime}_H\bra{\alpha}_H  
$$
\begin{equation}
+~ ~C_{\beta^\prime}^* (C_\beta C_\beta^* + C_\alpha^{*\,2})\ket{\alpha}_H\bra{\beta^\prime}_H ~+~ (C_\alpha^*+C_{\alpha}) C_{\beta^\prime}C_{\beta}^* ~\ket{\beta^\prime}_H\bra{\beta}_H~+~\dots  
\end{equation}
$$
$$
\begin{equation}
\Tr \varepsilon^3 ~=~C_{\alpha}^3 + C_{\alpha}^{*\,3}~+~3 (C_\alpha + C_\alpha^*) \sum_{\beta} C_\beta C_\beta^* +\dots=~\tilde{A}_{\alpha\alpha}^3 + \tilde{A}_{\alpha\alpha}^{*\,3}~+~3 (\tilde{A}_{\alpha\alpha} + \tilde{A}_{\alpha\alpha}^*) \sum_{\beta} C_\beta C_\beta^*~+~\dots
\end{equation}
The trace vanishes to order $e^6$ by unitarity.

Similarly, we obtain
\begin{equation}
\varepsilon^3\rho_0 ~=~ \left[(C_\alpha + 2C_\alpha^*) C_\beta C_\beta^* ~+~ C_\alpha^{*\,3}\right] ~\ket{\alpha}_H\bra{\alpha}_H~+~C_{\beta^\prime} \left[(C_\beta C_\beta^* + C_\alpha^2)~+~(C_\alpha^*+C_{\alpha})C_{\alpha}^*\right]~ \ket{\beta^\prime}_H\bra{\alpha}_H +\dots 
\end{equation}
$$
$$
\begin{equation}
\Tr(\varepsilon^3\rho_0)~=~C_{\alpha}^3 + C_{\alpha}^{*\,3}~+~ (C_\alpha + C_\alpha^*) \left(C_{\alpha}C_{\alpha}^*~+~2\sum_{\beta} C_\beta C_\beta^*\right)~+~\dots
\end{equation}
This trace does not contribute to the entanglement entropy at leading order.

Next we get
\begin{equation}
\varepsilon^2\rho_0\varepsilon\rho_0 ~=~ \left(C_{\alpha}~+~C_{\alpha}^*\right)\left(C_{\alpha}^{*\, 2}~+~C_\beta C_\beta^*\right)~\ket{\alpha}_H\bra{\alpha}_H 
~+~\left(C_{\alpha}~+~C_{\alpha}^*\right)\left(C_\beta C_\alpha~+~C_\beta C_{\alpha}^*\right) ~\ket{\beta}_H\bra{\alpha}_H +\dots 
\end{equation}
and
\begin{equation}
\Tr(\varepsilon^2\rho_0\varepsilon\rho_0)~=~\left(C_{\alpha}~+~C_{\alpha}^*\right)\left(C_{\alpha}^2~+~C_{\alpha}^{*\, 2}~+~C_{\alpha}C_{\alpha}^*~+~\sum_{\beta}C_\beta C_\beta^*\right)+\dots
\end{equation}
which vanishes to order $e^6$.

Finally to cubic order, we have
\begin{equation}
\varepsilon\rho_0\varepsilon\rho_0\varepsilon\rho_0~=~C_{\alpha}^*\left(C_{\alpha}~+~C_{\alpha}^*\right)^2~\ket{\alpha}_H\bra{\alpha}_H 
~+~C_{\beta}^*\left(C_{\alpha}~+~C_{\alpha}^*\right)^2~\ket{\beta}_H\bra{\alpha}_H +\dots 
\end{equation}
and
\begin{equation}
\Tr(\varepsilon\rho_0\varepsilon\rho_0\varepsilon\rho_0)=\left(C_{\alpha}~+~C_{\alpha}^*\right)^3+\dots
\end{equation}
with vanishing contributions to the entanglement entropy at leading order.

So 
\begin{equation}
\Tr\varepsilon,\,\, \Tr \varepsilon\rho_0 \varepsilon \rho_0 \varepsilon \rho_0, \,\, \Tr \varepsilon^2 \rho_0 \varepsilon \rho_0,\,\, \Tr\varepsilon^3\rho_0, \,\,\ \Tr\varepsilon^3, \,\, \Tr \varepsilon\rho_0\varepsilon\rho_0\,\, \rightarrow 0
\end{equation}
to order $e^6$ ($\Tr\varepsilon =0$ to all orders). The nonzero traces, which further simplify (by unitarity), are $\Tr \varepsilon \rho_0$,
$\Tr \varepsilon^2 $ and $\Tr \varepsilon^2 \rho_0$.

Recall from the previous section that
\begin{equation}
\bra{f_\beta}\ket{f_\alpha}= \left(\frac{\lambda}{E_d}\right)^{{\cal{B}}_{\beta\alpha}}=e^{{\cal{B}}_{\beta\alpha}\ln(\lambda/E_d)}=1+{\cal{B}}_{\beta\alpha}\ln(\lambda/E_d)+\dots \label{bracketf2}
\end{equation}
with ${\cal{B}_{\beta\alpha}}$ of order $e^2$ (given in \ref{B}), ${\cal{B}_{\alpha\alpha}}=0$, and
$$
{\tilde S}_{\beta\alpha}=
   ~\bra{f_\beta}\ket{f_\alpha}~
   ~\left(\frac{E_d}{\lambda}\right)^{2{\cal{B}}_{\beta\alpha}}~S_{\beta\alpha}~=~\left(\frac{E_d}{\lambda}\right)^{{\cal{B}}_{\beta\alpha}}~S_{\beta\alpha}
$$
$$
{\tilde S}_{\beta\gamma,\,\alpha}=
   ~\bra{f_\beta}\ket{f_\alpha}~
   ~\left(\frac{E_d}{\lambda}\right)^{2{\cal{B}}_{\beta\alpha}}~S_{\beta\gamma,\,\alpha}~=~\left(\frac{E_d}{\lambda}\right)^{{\cal{B}}_{\beta\alpha}}~S_{\beta\gamma,\,\alpha}~~~~{\rm{if}}~~~\omega_\gamma>E_d
$$
$$
{\tilde B}_{\beta\gamma,\,\alpha}~=~\frac{1}{(2V^5\omega_\gamma)^{1/2}}~F_{\beta\alpha}(\vec{q}_\gamma,\,\epsilon_r(\vec{q}_\gamma))~~~~{\rm{if}}~~~\omega_\gamma<E_d
$$
To leading order ($e^3$), the function $F_{\beta\alpha}(\vec{q}_\gamma,\,\epsilon_r(\vec{q}_\gamma))$ is smooth and nonsingular in the limits $\lambda,\,|\vec{q}_\gamma| \to 0$ (and of order the dressing scale $E_d$ upon suitably modifying the dressing function $f_\beta^\mu(\vec{q})$ to subleading order in $\vec{q}$ \cite{Choi2}). The volume factors are due to the relative normalization between box and continuum states -- see below. (Some energy factors of the initial and final electron states can be absorbed in the definition of $F_{\beta\alpha}(\gamma)$).

From these, it is easy to deduce the relations
\begin{equation}
{\tilde A}_{\beta\alpha}=A_{\beta\alpha}/\bra{f_\beta}\ket{f_\alpha}
\end{equation}
and
\begin{equation}
{\tilde B}_{\beta\gamma,\,\alpha}=B_{\beta\gamma,\,\alpha}/\bra{f_\beta}\ket{f_\alpha}~~~~{\rm{if}}~~~\omega_\gamma>E_d
\end{equation}
For the purposes of perturbation theory it will be more convenient to express the traces in terms of Fock basis amplitudes, which are easier to compute via Feynman diagrams. Since the dressed amplitudes are free of IR divergences order by order in perturbation theory, any logarithmic divergence in the IR cutoff $\lambda$ at the perturbative level can be attributed to the soft clouds of photons via the coherent state overlaps.  

We incorporate the results above and collect the terms contributing to the nonzero traces to order $e^6$. First we find
\begin{equation}
\Tr \varepsilon \rho_0 = A_{\alpha\alpha} + A^*_{\alpha\alpha} + A_{\alpha\alpha}A^*_{\alpha\alpha}~+~\left(\sum_{\omega_{\gamma}<E_d}\frac{1}{2V^5\omega_\gamma}F_{\alpha\alpha}(\gamma)F^*_{\alpha\alpha}(\gamma)~+~\sum_{E_d<\omega_{\gamma}<E}{B}_{\alpha\gamma,\,\alpha}{B}^*_{\alpha\gamma,\,\alpha}\right) \label{Finaltrace1}
\end{equation}
Let us discuss the two terms in the parentheses. The last term vanishes by energy conservation. In the continuum limit, the first term in the parentheses gives, up to $\lambda$ independent multiplicative factors,
\begin{equation}
\int_\lambda^{E_d}~\frac{d^3\vec{q}}{(2\pi)^32\omega_{\vec{q}}}~\sum_r|F_{\alpha\alpha}(\vec{q},\,\epsilon_r(\vec{q}))|^2
\end{equation}
Since the function $F_{\beta\alpha}(\gamma)$ is smooth in the limits $\lambda,\, |\vec{q}| \to 0$, the integral is of order $E_d^2$ (at most). Therefore, this contribution is suppressed and can be dropped. 

For the quadratic traces we get
$$
\Tr \varepsilon^2 = A_{\alpha\alpha}^2 + A_{\alpha\alpha}^{*\, 2}+2\sum_{\beta} \left(A_{\beta\alpha}A_{\beta\alpha}^* +\sum_{\omega_\gamma>E}B_{\beta\gamma,\,\alpha}B_{\beta\gamma,\,\alpha}^* \right)
$$
\begin{equation}
+2\sum_{\beta}\sum_{\omega_\gamma<{E_d}}\frac{1}{2V^3\omega_\gamma} \left[ A_{\beta\alpha}F_{\beta\alpha}^*(\gamma)\left(f_\alpha(\vec{q}_\gamma)-f_\beta(\vec{q}_\gamma)\right)\cdot \epsilon^*(\gamma)+A_{\beta\alpha}^*F_{\beta\alpha}(\gamma)\left(f_\alpha^*(\vec{q}_\gamma)-f_\beta^*(\vec{q}_\gamma)\right)\cdot \epsilon(\gamma)\right] \label{delta2}
\end{equation}
$$
$$
$$
\Tr \varepsilon^2 \rho_0 = A_{\alpha\alpha}^2 + A_{\alpha\alpha}^{*\, 2}+ A_{\alpha\alpha}A_{\alpha\alpha}^*+\sum_{\beta}\left(A_{\beta\alpha}A_{\beta\alpha}^*+\sum_{\omega_\gamma>E}B_{\beta\gamma,\,\alpha}B_{\beta\gamma,\,\alpha}^* \right)
$$
\begin{equation}
+\sum_{\beta}\sum_{\omega_\gamma<{E_d}}\frac{1}{2V^3\omega_\gamma} \left[ A_{\beta\alpha}F_{\beta\alpha}^*(\gamma)\left(f_\alpha(\vec{q}_\gamma)-f_\beta(\vec{q}_\gamma)\right)\cdot \epsilon^*(\gamma)+A_{\beta\alpha}^*F_{\beta\alpha}(\gamma)\left(f_\alpha^*(\vec{q}_\gamma)-f_\beta^*(\vec{q}_\gamma)\right)\cdot \epsilon(\gamma)\right] \label{delta2rho}
\end{equation}
The last lines in \ref{delta2} and \ref{delta2rho} arise due to the dressing. Notice that since $F_{\beta\alpha}(\gamma)$ is of order $e^3$ and the dressing function of order $e$, the amplitude $A_{\beta\alpha}$ must be computed at tree level, and so it does not exhibit any IR divergences as $\lambda \to 0$. In the continuum limit, these lines give rise to the following integral (up to smooth, nonsingular factors as $\lambda \to 0$ and volume factors):
\begin{equation}
\int_\lambda^{E_d}~\frac{d^3\vec{q}}{(2\pi)^32\omega_{\vec{q}}}~\sum_rF_{\beta\alpha}^*(\vec{q},\,\epsilon_r(\vec{q}))~\sum_{s\in\{\alpha,\,\beta\}}~\frac{e_s\eta_s~p_s\cdot\epsilon_r^*(\vec{q})}{p_s\cdot q} ~+~ h.c.
\end{equation}
Taking into account the measure of integration, the integrand is smooth in the $|\vec{q}| \to 0$ limit. So the integral is of order $E_d$. The last lines in \ref{delta2} and \ref{delta2rho} give negligible contributions to the entanglement entropy.

Now let us compute $\Tr (\rho_H)^2$ to order $e^6$. It is given by
$$
\Tr (\rho_H)^2 = \Tr \rho_0^2 + 2 \Tr \varepsilon \rho_0 + \Tr \varepsilon^2
$$
\begin{equation}
= 1 + 2(A_{\alpha\alpha} + A^*_{\alpha\alpha}) + (A_{\alpha\alpha} + A^*_{\alpha\alpha})^2 + 2\sum_{\beta} \left(A_{\beta\alpha}A_{\beta\alpha}^* +\sum_{\omega_\gamma>E}B_{\beta\gamma,\,\alpha}B_{\beta\gamma,\,\alpha}^* \right)
\end{equation}
%\be
%+2\sum_{\beta}\sum_{\omega_\gamma<{E_d}}\frac{1}{(2V\omega_\gamma)^{1/2}} \left[ A_{\beta\alpha}B_{\beta\gamma,\,\alpha}^*\left(f_\alpha(\vec{q}_%\gamma)-f_\beta(\vec{q}_\gamma)\right)\cdot \epsilon^*(\gamma)+A_{\beta\alpha}^*B_{\beta\gamma,\,\alpha}\left(f_\alpha^*(\vec{q}_\gamma)-f_\beta^%*(\vec{q}_\gamma)\right)\cdot \epsilon(\gamma)\right] 
%\ee
Using the unitarity relation, \ref{unitarity2}, this simplifies further to
\begin{equation}
\Tr (\rho_H)^2 = 1 - 2 \Delta
\end{equation}
where
\begin{equation}
\Delta=\sum_{\beta} \sum_{\omega_\gamma<E}B_{\beta\gamma,\,\alpha}B_{\beta\gamma,\,\alpha}^* 
\end{equation}
%\be
%- \sum_{\beta}\sum_{\omega_\gamma<{E_d}}\frac{1}{(2V\omega_\gamma)^{1/2}} \left[ A_{\beta\alpha}B_{\beta\gamma,\,\alpha}^*\left(f_\alpha(\vec{q}_%\gamma)-f_\beta(\vec{q}_\gamma)\right)\cdot \epsilon^*(\gamma)+A_{\beta\alpha}^*B_{\beta\gamma,\,\alpha}\left(f_\alpha^*(\vec{q}_\gamma)-f_\beta^%*(\vec{q}_\gamma)\right)\cdot \epsilon(\gamma)\right] 
%\ee
is an order $e^6$ quantity, which depends crucially on the {\it undressed} amplitude to emit a single soft photon with energy $\lambda<\omega_{\gamma}<E$.

Next we consider $\Tr (\rho_H)^m$ for $m\ge 3$. To order $e^6$, only two structures contribute: $\Tr \varepsilon \rho_0$ and $\Tr \varepsilon^2 \rho_0$ with both coefficients being equal to $m$. We get
\begin{equation}
\Tr (\rho_H)^m = 1 + m \Tr \varepsilon \rho_0 + m \Tr \varepsilon^2 \rho_0
\end{equation}
Using \ref{delta2rho} and \ref{unitarity2}, it is easy to see that
\begin{equation}
\Tr (\rho_H)^m = 1 - m \Delta
\end{equation}
This result is in accordance with the fact that $\rho_H$ has one large eigenvalue, which, to order $e^6$, is equal to $1-\Delta$. All of the rest nonvanishing eigenvalues are of order $e^6$ (or higher), and their sum is equal to $\Delta$. This sum sets the behavior of the leading entanglement entropy.

\subsection{Entanglement entropy}
\noindent

We proceed now to compute the Renyi entropies to leading order in perturbation theory.
For any $m\ge 1$, we obtain
\begin{equation}
S_{m+1} = -\frac{1}{m}\log\left[1 - (m+1) \Delta\right]=\frac{m+1}{m}\Delta
\end{equation}
Using \ref{entR},
%\begin{equation}
%S_{ent}=\sum_{n=1}^{\infty}\sum_{m=0}^{n}\frac{(n-1)!}{(n-m)!~m!}~(-1)^m~e^{-mS_{m+1}} 
%\end{equation}
the perturbative result for the Renyi entropies and the identity
$$
\sum_{m=0}^n (^{\, n}_{\, m}) (-1)^m =0
$$
we obtain for the leading entanglement entropy
\begin{equation}
S_{ent}=-\Delta~\ln{e^6}
\end{equation}
(The leading entanglement entropy is of order $e^6\ln{e^6}$).

Now $\Delta$ is singular in the limit $\lambda \to 0$. Let us examine the singular part. We have
\begin{equation}
\Delta_{sing}=\sum_{\beta} \sum_{\omega_\gamma<{E_d}}B_{\beta\gamma,\,\alpha}B_{\beta\gamma,\,\alpha}^* 
\end{equation}
%\be
%- \sum_{\beta}\sum_{\omega_\gamma<{E_d}}\frac{1}{(2V\omega_\gamma)^{1/2}} \left[ A_{\beta\alpha}B_{\beta\gamma,\,\alpha}^*\left(f_\alpha(\vec{q}_%\gamma)-f_\beta(\vec{q}_\gamma)\right)\cdot \epsilon^*(\gamma)+A_{\beta\alpha}^*B_{\beta\gamma,\,\alpha}\left(f_\alpha^*(\vec{q}_\gamma)-f_\beta^%*(\vec{q}_\gamma)\right)\cdot \epsilon(\gamma)\right] 
%\ee
%and
%\be
%\Delta_{sing}=\sum_{\beta} \sum_{\omega_\gamma<{E_d}}B_{\beta\gamma,\,\alpha}B_{\beta\gamma,\,\alpha}^* 
%\ee
Using soft photon theorems, we find
\begin{equation}
\Delta_{sing}=\sum_{\beta} e^2 (A_{\beta\alpha}A_{\beta\alpha}^*) \left[\sum_{\omega_\gamma<{E_d}}\frac{1}{(2V\omega_\gamma)}
\sum_{ss^\prime\in\{\alpha,\,\beta\}} \eta_s \eta_{s^\prime} ~\frac{p_sp_{s^\prime}}{(p_sq_\gamma)(p_{s^\prime}q_\gamma)}\right] \label{DeltaSingular}
\end{equation}
where the undressed amplitude $A_{\beta\alpha}$ is computed at tree level.

{\it The same result is obtained in the Fock basis case, in the absence of dressing}. As we have seen, the dressing adds
negligible contributions of order $E_d$ to the entanglement entropy and does not alleviate logarithmic singularities at the leading perturbative level -- see \ref{Finaltrace1}, \ref{delta2} and \ref{delta2rho} and the discussions around them. It would be
interesting to verify this result to all orders in perturbation theory. 
%$S_{ent}= \Delta$, where
%\be
%\Delta= \sum_{\beta} \sum_{\omega_\gamma<E}B_{\beta\gamma,\,\alpha}B_{\beta\gamma,\,\alpha}^* 
%\ee

\subsection{Continuum limit}\label{continuum}
\noindent
To take the continuum limit, recall that a box single particle state (which is normalizable) is related to a continuum single particle state
(which is $\delta$-function normalizable) by the factor
\begin{equation}
\ket{\vec{p}}_{Box} \to \frac{1}{(2E_{\vec{p}}~V)^{1/2}}\ket{\vec{p}}
\end{equation}
So in the continuum limit we obtain for the singular part of the entanglement entropy
$$
S_{ent,\,sing} = -\frac{e^2}{2\,V^2} \int\frac{d^3\vec{p}_k}{(2\pi)^32E_k}~\int\frac{d^3\vec{p}_l}{(2\pi)^32E_l}~\frac{\ln{e^6}}{2E_i~2E_j}~\left|i{\cal{M}}_{kl}^{ij}\right|^2~\left[(2\pi)^4\delta^4(p_k + p_l  - p_i -p_j)\right]^2
$$
\begin{equation}
\times ~\int_{\lambda}^{E_d}\frac{d^3\vec{q}}{(2\pi)^32\omega_{\vec{q}}}~\sum_{ss^\prime\in\{i,j,k,l\}} \eta_s \eta_{s^\prime} ~\frac{p_sp_{s^\prime}}{(p_sq)(p_{s^\prime}q)} \label{EntSingular}
\end{equation}
where $i{\cal{M}}_{kl}^{ij}$ is the invariant amplitude for the process $e_i + e_j \to e_k + e_l$ (Moller scattering), given in terms of tree level Feynman diagrams. Integration over $\vec{p}_l$ imposes momentum conservation, $\vec{p}_l=\vec{p}_i+\vec{p}_j-\vec{p}_k$, and yields an additional volume factor in the numerator ($(2\pi)^3 \delta^3(0)=V$). Integrating in addition over the soft photon momentum yields the logarithmically divergent factor 
\begin{equation}
S_{ent,\,sing} = -\frac{1}{V}~\ln\left(\frac{E_d}{\lambda}\right)~ \int\frac{d^3\vec{p}_k}{(2\pi)^32E_k}~\frac{\ln{e^6}}{8E_iE_jE_l}~\left|i{\cal{M}}_{kl}^{ij}\right|^2~{\cal{B}}_{kl,\,ij}~\left[(2\pi)\delta(E_k + E_l  - E_i -E_j)\right]^2
\end{equation}
where ${\cal{B}}_{kl,\,ij}={\cal{B}}_{\beta\alpha}$ is given by \ref{B}.

We let the incoming electrons have opposite momenta along the $z$-axis, $\vec{p}_i=-\vec{p}_j=p_0 \hat{z}$, working in the center of mass frame. Without loss of generality we take $p_0$ to be positive. The center of mass energy is $E_{cm} =2E_i= 2\sqrt{p_0^2 + m^2}$.

Thus in this frame, we may set $\vec{p}_k=-\vec{p}_l=p^{\prime} \hat{k}$ and $E_k=E_l =\sqrt{p^{\prime\, 2}+m^2}$. Integration over the magnitude of $\vec{p}_k$ imposes energy conservation, $|p^{\prime}|=p_0$ (or $E_k = E_l = E_{cm}/2$), and yields a factor of $2\pi \delta (E_i - E_i)=T$, with $T$ the timescale of the experiment. We finally obtain
\begin{equation}
S_{ent,\, sing} = -\frac{T\,v_{rel}}{16\,V}~\ln\left(\frac{E_d}{\lambda}\right)~ \int\frac{d^2\hat{k}}{(2\pi)^2}~\frac{\ln{e^6}}{E_{cm}^2}\left|i{\cal{M}}_{kl}^{ij}\right|^2~{\cal{B}}_{kl,\,ij} \label{Entd}
\end{equation}
where $v_{rel}=2p_0/E_i=4p_0/E_{cm}$ is the relative velocity of the particles. 

Now $v_{rel}/V$ is the flux of particle $j$ with respect to particle $i$ (and vice versa). We define the entanglement entropy per flux per unit time, $s_{ent}$, to find
\begin{equation}
s_{ent,\, sing} = -\frac{1}{16}~\ln\left(\frac{E_d}{\lambda}\right)~ \int\frac{d^2\hat{k}}{(2\pi)^2}~\frac{\ln{e^6}}{E_{cm}^2}\left|i{\cal{M}}_{kl}^{ij}\right|^2~{\cal{B}}_{kl,\,ij}
\end{equation}
Notice that the integrand is a function of the scattering angle $\theta$ ($\cos\theta = \hat{k}\cdot \hat{z}$). For slowly moving particles, the Moller amplitude squared (averaged over spin polarizations) scales as $\left|i{\cal{M}}_{kl}^{ij}\right|^2 \sim e^4 m^4/p_0^4\sin^4\theta$. Likewise ${\cal{B}}_{kl,\,ij}$ scales as $\sin^2\theta$. So the integrand diverges for forward ($\theta=0$) and backward ($\theta=\pi$) scattering. However, scattering at $\theta = 0$ or $\theta = \pi$ cannot be distinguished from no scattering. This introduces an effective lower and upper cutoff $\theta_0\le \theta \le \pi -\theta_0$ on the scattering angle, which regularizes the above integral.

Recall that to leading order only single photon particle states contribute to the traces and the entanglement entropy -- see \ref{DeltaSingular}. The dimensionality of the subspace of single photon states with frequency less than $E_d$ scales as $D\sim (E_dL)^3$, where $L$ is the size of the box (and becomes infinite in the continuum limit). In fact the entanglement entropy between the soft and the hard particles cannot exceed $\log D$. Taking $\lambda$ to be of order $1/L$, we see that the dominant contribution to the entanglement entropy \ref{Entd} is a fraction of the maximum possible value.

Thus the perturbative calculation of the entanglement entropy associated with tracing over all soft photons in ${\cal{H}}_S$ breaks down in the strict $\lambda \to 0$ limit. The logarithmic divergences in $\lambda$ do not cancel order by order in perturbation theory. One may wonder if the entanglement entropy to all orders is finite in the continuum limit, since the reduced density matrix is dominated by the diagonal elements which are free of any IR divergences. Notice that the diagonal element \ref{Ddiagonal} scales inversely proportional with a power of the volume $V$ in the continuum limit. So the entanglement entropy per flux per unit time is expected to diverge logarithmically in the volume: $s_{ent} \sim \log V$. At the intuitive level, this behavior can be understood as follows. The density matrix becomes very incoherent in this limit. We expect the number of its nonzero eigenvalues to be of order the dimensionality $D_S$ of ${\cal{H}}_S$ \footnote{Suppose a quantum system consists of two subsystems $A$ and $B$, with dimensionalities $D_B > D_A$. Let the whole system be in a pure state. Then the density matrices describing the two subsystems have equal non-zero eigenvalues. When maximal disorder is reached, the number of non-zero eigenvalues attains its maximal possible value, set by the smaller dimensionality $D_A$. Moreover, the non-zero eigenvalues become equal to each other, and so equal to $1/D_A$. The entanglement entropy is equal to $\log D_A$.}, and each to scale with $1/D_S \sim 1/V$. The entanglement entropy scales with $\log V$.  

We emphasize that the singular part of the entanglement entropy, \ref{DeltaSingular} (and \ref{EntSingular} in the continuum limit), does not depend on the details of the Faddeev-Kulish dressing. As we have already explained, precisely the same expression is obtained in the Fock basis computation, where the initial state is taken to be undressed. The structure of the expression is suggestive of a universal applicability to generic scattering processes. Namely, the leading entanglement entropy is given as a sum over transition probabilities (for the initial state $\alpha$ to scatter to a final hard state $\beta$), with each probability weighted by a soft photon factor. Integration over the soft photon momentum gives rise to the logarithmic singularity in $\lambda$. It would be interesting to see if and how higher order corrections in the electron coupling modify this structure.

It is interesting to contrast our findings concerning the entanglement between soft and hard degrees of freedom after scattering with other examples of entanglement in quantum field theory, such as the entanglement between the local degrees of freedom associated with a region of space and the degrees of freedom of its complement -- See e.g. \cite{Rangamani}. The entanglement entropy in this case is UV divergent (unlike the case studied in this work, where the divergence is infrared in nature), with the divergences arising from local effects. As the case at hand, the coefficients of the singular terms are universal and contain physical information. For example, the coefficient of the leading quadratic term in the UV cutoff is proportional to the area of the boundary of the region and the number of degrees of freedom of the field theory. Likewise the coefficient of the logarithmic singular term scales with the number of degrees of freedom and depends on the shape of the boundary via an integral of $K^2$, where $K$ is the trace of the second fundamental form of the induced metric on the boundary surface \cite{MaldacenaENT,Solodukhin}.

The coefficient of the IR logarithmic singularity in our case also contains physical information. The soft photon factor in the last line of \ref{EntSingular} gives
${\cal{B}}_{kl,\,ij}~\ln\left(E_d/\lambda\right)$,
with ${\cal{B}}_{kl,\,ij}$ given by \ref{B}. In terms of Mandelstam variables we obtain
\begin{equation}
{\cal{B}}_{kl,\,ij}~=~\frac{e^2}{4\pi^2}~\left[~\frac{1-\frac{2m^2}{t}}{\sqrt{1-\frac{4m^2}{t}}}\ln\left(\frac{1-\frac{2m^2}{t}+\sqrt{1-\frac{4m^2}{t}}}{1-\frac{2m^2}{t}-\sqrt{1-\frac{4m^2}{t}}}\right)~+~\left(t \leftrightarrow u\right)~-~\left(t \leftrightarrow s\right)~-~2~\right]
\end{equation}
Let us consider high energy and small angle scattering, keeping $t$ to be large and fixed. In this limit, we get
\begin{equation}  
{\cal{B}}_{kl,\,ij}~\simeq~\frac{e^2}{4\pi^2}\ln\left(\frac{|t|}{4m^2}\right)
\end{equation}
and so the soft photon factor gives rise to a double log contribution. The coefficient then becomes equal to the cusp anomalous dimension in QED, $\Gamma({\varphi},\;\alpha)$ via the relation $|t|=2m^2 (\cosh{\varphi} - 1)$, controlling the vacuum expectation value of a Wilson loop with a cusp of angle $\varphi$ -- See e.g. \cite{Korchemsky} and \cite{Sever2} for discussions.

\subsection{Soft radiation and entanglement}
\noindent
We proceed now to study the reduced density matrix obtained by tracing over soft radiation photons with frequencies $E_d<\omega_\gamma<E$, as advocated also in \cite{Gomez2}. This tracing is motivated by the fact that starting with initial dressed states, the amplitude to emit a photon with energy below the dressing scale $E_d$ is suppressed. The density matrix takes the following form
$$
\rho_{asym}=\ket{\alpha}_d\bra{\alpha}_d + \left(\tilde{A}_{\beta\alpha}\ket{\beta}_d~+~\sum_{\omega_\gamma <E_d, \,\omega_\gamma>E}\tilde{B}_{\beta\gamma,\,\alpha}\ket{\beta\gamma}_d~+~\dots\right)\bra{\alpha}_d
$$
$$
+\ket{\alpha}_d\left(\tilde{A}^*_{\beta^\prime\alpha}\bra{\beta^\prime}_d~+~\sum_{\omega_{\gamma^\prime}<E_d,\,\,\omega_{\gamma^\prime}>E}\tilde{B}^*_{\beta^\prime\gamma^\prime,\,\alpha}\bra{\beta^\prime\gamma^\prime}_d~+~\dots\right)
$$
$$
+\left(\tilde{A}_{\beta\alpha}\tilde{A}^*_{\beta^\prime\alpha}~+~\sum_{E_d<\omega_\gamma<E}\tilde{B}_{\beta\gamma,\,\alpha}\tilde{B}^*_{\beta^\prime\gamma,\,\alpha}\right)\ket{\beta}_d\bra{\beta^\prime}_d~+~\sum_{\omega_\gamma,\omega_{\gamma^\prime}<E_d,\,\,\omega_\gamma,\omega_{\gamma^\prime}>E}\tilde{B}_{\beta\gamma,\,\alpha}\tilde{B}^*_{\beta^\prime\gamma^\prime,\,\alpha} \ket{\beta\gamma}_d\bra{\beta^\prime\gamma^\prime}_d 
$$
\begin{equation}
+ \sum_{\omega_\gamma<E_d,\,\omega_\gamma>E}\tilde{B}_{\beta\gamma,\,\alpha}\tilde{A}^*_{\beta^\prime\alpha}\ket{\beta\gamma}_d\bra{\beta^\prime}_d~+~\sum_{\omega_{\gamma^\prime}<E_d,\,\omega_{\gamma^\prime}>E}\tilde{B}^*_{\beta^\prime\gamma^\prime,\,\alpha}\tilde{A}_{\beta\alpha}\ket{\beta}_d\bra{\beta^\prime\gamma^\prime}_d  ~+~ \dots
\end{equation}

This density matrix is an operator acting on the space of asymptotic states. The matrix elements are given exclusively in terms of dressed amplitudes -- the overlaps $\bra{f_{\beta^\prime}}\ket{f_\beta}$ are absent. Thus, no IR divergences appear in the $\lambda \to 0$ limit at any finite order in perturbation theory. Moreover, the off-diagonal elements remain nonvanishing in the $\lambda \to 0$ limit (as compared with the diagonal elements), to all orders in perturbation theory. The density matrix does not exhibit decoherence. Since the dressed amplitude $\tilde{B}_{\beta\gamma,\,\alpha}$ to emit a photon of energy less than $E_d$ is suppressed, the contributions of various sums over photon frequencies smaller than $E_d$ can be neglected.  

Now let us compute the entanglement entropy to leading order in perturbation theory. As before we set $\varepsilon=\rho_{asym}-\rho_0$, where now $\rho_0=\ket{\alpha}_d\bra{\alpha}_d$, $\rho_0^2=\rho_0$ and $\Tr \varepsilon =0$ by unitarity. At leading order ($e^6$), the only nonvanishing traces are
\begin{equation}
    \Tr \varepsilon\rho_0 ~=~ \tilde{A}_{\alpha\alpha} ~+~ \tilde{A}_{\alpha\alpha}^* ~+~ \tilde{A}_{\alpha\alpha}\tilde{A}_{\alpha\alpha}^* ~+~ \sum_{E_d<\omega_\gamma<E}\tilde{B}_{\alpha\gamma,\,\alpha}\tilde{B}^*_{\alpha\gamma,\,\alpha}
\end{equation}
$$
$$
\begin{equation}
    \Tr \varepsilon^2 ~=~ \tilde{A}_{\alpha\alpha}^2 ~+~ \tilde{A}_{\alpha\alpha}^{*\,2}~+~2~\sum_{\beta}\left[(1 + \tilde{A}_{\alpha\alpha} +  \tilde{A}_{\alpha\alpha}^*)\tilde{A}_{\beta\alpha}\tilde{A}_{\beta\alpha}^* ~+~ \sum_{\omega_\gamma<E_d,\,\omega_{\gamma}>E}\tilde{B}_{\beta\gamma,\,\alpha}\tilde{B}^*_{\beta\gamma,\,\alpha}\right]
\end{equation}
$$
$$
$$
\Tr \varepsilon^2\rho_0 ~=~\tilde{A}_{\alpha\alpha}^2~+~\tilde{A}_{\alpha\alpha}^{*\,2}~+~\tilde{A}_{\alpha\alpha}\tilde{A}_{\alpha\alpha}^*(1+\tilde{A}_{\alpha\alpha}+\tilde{A}_{\alpha\alpha}^*)
$$
\begin{equation}
+~\sum_{\beta}\left[(1+ \tilde{A}_{\alpha\alpha} +  \tilde{A}_{\alpha\alpha}^*) \tilde{A}_{\beta\alpha} \tilde{A}_{\beta\alpha}^* ~+~ \sum_{\omega_\gamma<E_d,\,\omega_{\gamma}>E}\tilde{B}_{\beta\gamma,\,\alpha}\tilde{B}^*_{\beta\gamma,\,\alpha}\right]
\end{equation} 
These can be further simplified using \ref{unitarity2}. Also to order $e^6$
$$
\sum_{E_d<\omega_{\gamma}<E}\tilde{B}_{\alpha\gamma,\,\alpha}\tilde{B}^*_{\alpha\gamma,\,\alpha}~=~\sum_{E_d<\omega_{\gamma}<E}B_{\alpha\gamma,\,\alpha}B^*_{\alpha\gamma,\,\alpha}=0
$$
the latter vanishing by energy conservation.

Thus 
$$
\Tr(\rho_H)^2 ~=~ 1~+~2\Tr\varepsilon\rho_0~+~\Tr\varepsilon^2~=~1~+~2~(\tilde{A}_{\alpha\alpha} + \tilde{A}_{\alpha\alpha})
$$
\begin{equation}
 +~2~\sum_{\beta}\left( \tilde{A}_{\beta\alpha} \tilde{A}_{\beta\alpha}^* ~+~ \sum_{\omega_\gamma<E_d,\,\omega_{\gamma}>E}\tilde{B}_{\beta\gamma,\,\alpha}\tilde{B}^*_{\beta\gamma,\,\alpha}\right)~=~1~-~2~\sum_{\beta}\sum_{E_d<\omega_{\gamma}<E}\tilde{B}_{\beta\gamma,\,\alpha}\tilde{B}^*_{\beta\gamma,\,\alpha}
\end{equation}
We used \ref{unitarity2} and we dropped terms of order $e^6$. Likewise we can show
\begin{equation}
\Tr(\rho_H)^m ~=~ 1~+~m\Tr\varepsilon\rho_0~+~\Tr\varepsilon^2\rho_0~=~1~-~m~\sum_{\beta}\sum_{E_d<\omega_{\gamma}<E}\tilde{B}_{\beta\gamma,\,\alpha}\tilde{B}^*_{\beta\gamma,\,\alpha}
\end{equation}

The Renyi entropies and the entanglement entropy are given by
\begin{equation}
S_{m+1} =\frac{m+1}{m}\sum_{\beta}\sum_{E_d<\omega_{\gamma}<E}\tilde{B}_{\beta\gamma,\,\alpha}\tilde{B}^*_{\beta\gamma,\,\alpha},\,\,\, m\ge 1
\end{equation}  
and
\begin{equation}
S_{ent}= -\ln{e^6}~\sum_{\beta}\sum_{E_d<\omega_{\gamma}<E}\tilde{B}_{\beta\gamma,\,\alpha}\tilde{B}^*_{\beta\gamma,\,\alpha} 
\end{equation}
To order $e^6\ln{e^6}$, this is given by
\begin{equation}
S_{ent}= -\ln{e^6}~\sum_{\beta}\sum_{E_d<\omega_{\gamma}<E}{B}_{\beta\gamma,\,\alpha}{B}^*_{\beta\gamma,\,\alpha} 
\end{equation}
where the dressing scale $E_d$ provides the lower cutoff. In particular, $E_d$ is kept finite in the continuum, $\lambda \to 0$ limit, and so the leading perturbative entanglement entropy is finite.

Letting the energy scale $E$ to be sufficiently small and repeating steps as in the previous section, we obtain for the entanglement entropy per unit flux per unit time in the continuum limit:
\begin{equation}
s_{ent} = -\frac{1}{16}~\ln\left(\frac{E}{E_d}\right)~ \int\frac{d^2\hat{k}}{(2\pi)^2}~\frac{\ln{e^6}}{E_{cm}^2}\left|i{\cal{M}}_{kl}^{ij}\right|^2~{\cal{B}}_{kl,\,ij} \label{entfinite}
\end{equation}
This quantity is finite in the $\lambda \to 0$ limit. Notice that as $E \to E_d$, the entanglement entropy becomes vanishingly small. In particular the radiated soft photons carry little information. We should emphasize that expression \ref{entfinite} arises when the IR scales $E$ and $E_d$ are sufficiently small. In general the expression for the entanglement entropy per unit flux per unit time in the continuum limit will be more complicated, depending on the scales $E_d$ and $E$.

%%%%%%%%%%%%%%%%%%%%%%%%%%%%%%%%%%%%%%%%%%%%%%%%%%%%%%%%%%%%%%%%%%%%%%%%%%%%%%%%%%
\section{Conclusions}\label{s5}
%%%%%%%%%%%%%%%%%%%%%%%%%%%%%%%%%%%%%%%%%%%%%%%%%%
\noindent

In this paper we studied the entanglement between the hard and soft particles produced during a typical scattering process of Faddeev-Kulish electrons in QED. Tracing over the entire spectrum of soft photons leads to decoherence and infrared divergences in the perturbative expansion for the entanglement entropy. To leading order, the entanglement entropy is set by the conventional Fock basis amplitude squared for real single soft photon emission, leading to a logarithmic infrared divergence when integrated over the soft momentum. The same result is obtained in a Fock basis computation, where the initial state consists of two bare electrons. In particular, the singular part of the entanglement entropy does not depend on the details of the Faddeev-Kulish dressing. The expression is suggestive for a universal applicability to generic scattering processes. For the case of Faddeev-Kulish electrons though the divergence can be traced in the overlap of the coherent states describing the soft photon clouds that accompany the asymptotic charged particles. Thus there is strong entanglement between the final state hard charged particles and the photons in the clouds. 

By suitably modifying the dressing function to subleading order in the soft momentum, one can show that the Faddeev-Kulish
amplitudes for the emission of soft photons with energies less than $E_d$, the characteristic energy of photons in the clouds, are suppressed (of order $E_d$), at least at tree level \cite{Choi2}. This suggests that the soft part of the emitted radiation consists of low energy photons with energy greater than the dressing scale $E_d$. Taking a partial trace over these soft radiative photons produces a well defined density matrix, free of any infrared divergences order by order in perturbation theory. The reduced density matrix is now an operator acting on the space of asymptotic states, and does not exhibit large amount of decoherence \cite{Gomez2}. The entanglement entropy is free of any infrared divergences at any order in the perturbative expansion. As the energy set by the resolution of the detector approaches the effective cutoff scale $E_d$, provided by the dressing, the leading entanglement entropy becomes vanishingly small, suggesting that a small amount of information is carried by the soft radiated photons. It would be interesting to see if the suppression of the Faddeev-Kulish amplitudes for the emission of soft photons with energies less than $E_d$ persists at the one loop level, since then logarithmic corrections in the soft photon frequency appear. One would need to consider higher order corrections to the dressing function to implement this task.      

It would also be interesting to investigate the applicability of our results to the case of gravity. At least the perturbative analysis in this work suggests strong correlations between the hard particles produced in a scattering process and the soft gravitons present in the clouds accompanying them. Conservation laws associated with large gauge transformations (supertranslations and superrotations) require the hard Hawking quanta produced during the process of formation/evaporation of a black hole, to be accompanied by clouds of soft gravitons and photons \cite{StromingerBHinfo,HPS}. Despite the entanglement between these hard and soft degrees of freedom, it is difficult to see how black hole evaporation would result in a pure state of properly dressed, asymptotic particles, without invoking correlations between early and late time Hawking quanta \cite{Page}. Arguments suggesting the decoupling of soft variables from the hard dynamics seem to support this point of view \cite{Porrati2}.

%%%%%%%%%%%%%%%%%%%%%%%%%%%%%%%%%%%%%%%%%%%%%%%%%%%%%%%%%%%%%%%%%%%%%%%%%%%%%%%%%%
\section*{Acknowledgements} We thank C. Bachas and E. Kiritsis for useful discussions. N.T. wishes to thank the ITCP and the Department of Physics of the University of Crete where parts of this work were done for hospitality. 
%%%%%%%%%%%%%%%%%%%%%%%%%%%%%%%%%%%%%%%%%%%%%%%%%%
\noindent

%%%%%%%%%%%%%%%%%%%%%%%%%%%%%%%%%%%%%%%%%%%%%%%%%%%%%%%%%%%%%%%%%%%%%%%%%%%%%%%%%

\bigskip
\appendix
\labelformat{section}{Appendix #1} 
%%%%%%%%%%%%%%%%%%%%%%%%%%%%%%%%%%%%%%%%%%%%%%%%%%%%%%%%%%%%%%%%%%%%%%%%%%%%%%%%%
%%%%%%%%
%%%%%%%%%%%%%%%%%%
%%%%%%%%%%%%%%%%%%%%%%%%%%%%%%%%%%%%%%%%%%%%%%%%%
\section{Notations and conventions}\label{A1}
%%%%%%%%%%
Throughout we employ the Lorenz gauge, $\partial_{\mu}A^{\mu}=0$, with the free electromagnetic gauge field satisfying $\Box A^{\mu} =0$. We work with a mostly plus signature metric.

At the quantum level we expand the gauge field in terms of creation and annihilation operators
\begin{equation}
A^{\mu}(x)~=~\int \frac{d^3\vec{k}}{(2\pi)^3}~\frac{1}{\sqrt{\omega_{\vec{k}}}}~\sum_r~\epsilon^{\mu}_r(\vec{k})~a_r(\vec{k})~e^{ikx}~+~\epsilon^{\mu\, *}_r(\vec{k})~a_r^{\dagger}(\vec{k})~e^{-ikx}
\end{equation}
The four polarization vectors $\epsilon_r^{\mu}(\vec{k})$, $r=0,\dots,3$, satisfy the following orthonormality and completeness relations
\begin{equation}
\epsilon_{r\,\mu}(\vec{k})\epsilon_s^{\mu\,*}(\vec{k})=~\zeta_r~\delta_{rs},\,\,\,\,\, \sum_r~\zeta_r~\epsilon_r^{\mu}(\vec{k})\epsilon_r^{\nu\,*}(\vec{k})=\eta^{\mu\nu}   
\end{equation}
where $\zeta_0 = -1$ and $\zeta_1=\zeta_2=\zeta_3=1$.
The photon creation and annihilation operators satisfy the following commutation relations
\begin{equation}
[a_r(\vec{p}), a_{r^\prime}^\dagger(\vec{p}^\prime)]=(2\pi)^3~\zeta_r~\delta_{rr^\prime}~\delta^3(\vec{p}-\vec{p}^\prime) 
\end{equation}
In the quantum theory we impose the Gupta-Bleuler condition:
\begin{equation}
\left[a_0(\vec{k})-a_3(\vec{k})\right]\ket{\Psi}=0
\end{equation}
Finally the electron/positron creation and annihilation operators satisfy the following anticommutation relations  
\begin{equation}
\{b_s(\vec{p}), b_{s^\prime}^\dagger(\vec{p}^\prime)\}=(2\pi)^3\delta_{ss^\prime}\delta^3(\vec{p}-\vec{p}^\prime),\,\,\{d_s(\vec{p}), d_{s^\prime}^\dagger(\vec{p}^\prime)\}=(2\pi)^3\delta_{ss^\prime}\delta^3(\vec{p}-\vec{p}^\prime)
\end{equation}

%%%%%%%%%%%%%%%%%%%%%%%%%%%%%%%%%%%%%%%%%%%%%%%%%%%%%%%%%%%%%%%%%%%%%%%%%%%%%%%%%%%%%%%%%%%
\section{Multi electron/positron dressed states}\label{A2}

It will be useful to compute the normalization factor ${\cal{N}}_\alpha$ for the photon coherent state associated with the state $\alpha=\{e_i,\, \vec{p}_i,\, s_i\}$. It takes the form of \ref{normalization} with the exponent replaced by
$$
\int_\lambda^{E_d}\frac{d^3\vec{q}}{2(2\pi)^3}~\frac{1}{2\omega_{\vec{q}}}~f_\alpha^{\mu}(\vec{q})f^{*}_{\alpha\,\mu}(\vec{q}) = \int_\lambda^{E_d}\frac{d^3\vec{q}}{2(2\pi)^3}~\frac{1}{2\omega_{\vec{q}}}~\sum_{ij \in \alpha} e_ie_j\left(\frac{p_ip_j}{(p_iq)(p_jq)}-\frac{cp_i}{p_iq}-\frac{cp_j}{p_jq}\right)
$$
$$
= \frac{1}{4} \ln{\left(\frac{E_d}{\lambda}\right)}~\sum_{ij \in \alpha} e_i e_j~\int \frac{d^2\hat{q}}{(2\pi)^3}~\left(\frac{p_ip_j}{(p_i^0 - \vec{p}_i\cdot \hat{q})(p_j^0 - \vec{p}_j\cdot \hat{q})}+\frac{p_i^0 +\vec{p}_i\cdot \hat{q} }{2(p_i^0 - \vec{p}_i\cdot \hat{q})}+\frac{p_j^0 +\vec{p}_j\cdot \hat{q} }{2(p_j^0 - \vec{p}_j\cdot \hat{q})}\right)
$$
\begin{equation}
=\frac{1}{8\pi^2}~\ln{\left(\frac{E_d}{\lambda}\right)}~\sum_{ij \in \alpha}~ e_ie_j ~ I(\vec{v}_i,\, \vec{v}_j)
\end{equation}
where the kinematical factor is given by
\begin{equation}
I(\vec{v}_i,\, \vec{v}_j)=\frac{1}{2v_i}~ \ln \left(\frac{1+ v_i}{1-v_i}\right)~+~i \to j~-~\frac{1}{2v_{ij}}~ \ln \left(\frac{1+ v_{ij}}{1-v_{ij}}\right)~-~1
\end{equation}
Here $v_{ij}$ is (the magnitude of) the relative velocity of particle $j$ with respect to $i$:
\begin{equation}
v_{ij}=\left[1-\frac{m_i^2\,m_j^2}{(p_i\cdot p_j)^2}\right]^{1/2}
\end{equation}
Notice that the diagonal terms $i=j$ reduce to expression \ref{kinematical1}. 
Setting
\begin{equation}
{\cal{A}}_{\alpha}=\frac{1}{8\pi^2}~\sum_{ij \in \alpha}~ e_ie_j ~ I(\vec{v}_i,\, \vec{v}_j)
\end{equation}
we get
\begin{equation}
{\cal{N}}_\alpha=\left(\frac{\lambda}{E_d}\right)^{{\cal{A}}_{\alpha}}
\end{equation}
The exponent ${\cal{A}}_{\alpha}$ is non-negative, and so generically ${\cal{N}}_\alpha$ vanishes in the limit $\lambda \to 0$.

Another useful quantity to consider is the overlap between coherent photon states, corresponding to generic charged states $\alpha=\{e_i,\, \vec{p}_i,\,s_i\}$ and $\beta=\{e_i^{\prime}, \,\vec{p}_i^{\,\prime},\,s_i^{\,\prime}\}$:
\begin{equation}
\langle f_\beta|f_\alpha\rangle = {\cal{N}}_\beta\, {\cal{N}}_\alpha\, e^{\int_\lambda^{E_d}\frac{d^3\vec{q}}{(2\pi)^3}~\frac{1}{2\omega_{\vec{q}}}~f_\alpha^{\mu}(\vec{q})f^{*}_{\beta\,\mu}(\vec{q}) } 
\end{equation}
The new exponent
\begin{equation}
\int_\lambda^{E_d}\frac{d^3\vec{q}}{(2\pi)^3}~\frac{1}{2\omega_{\vec{q}}}~f_\alpha^{\mu}(\vec{q})f^{*}_{\beta\,\mu}(\vec{q})
\end{equation}
is calculated to be
\begin{equation}
\int_\lambda^{E_d}\frac{d^3\vec{q}}{(2\pi)^3}~\frac{1}{2\omega_{\vec{q}}}~\sum_{i \in \alpha}\sum_{j \in \beta}~e_ie_j~\left(\frac{p_ip_j}{(p_iq)(p_jq)}~-~\frac{cp_i}{p_iq}~-~\frac{cp_j}{p_jq}\right)=\frac{1}{4\pi^2}~\ln{\left(\frac{E_d}{\lambda}\right)}~\sum_{i \in \alpha}\sum_{j \in \beta}~e_ie_j~I(\vec{v}_i,\, \vec{v}_j)
\end{equation}
Setting
\begin{equation}
{\cal{A}}_{\beta\alpha}=\frac{1}{4\pi^2}~\sum_{i \in \alpha}\sum_{j \in \beta}~e_ie_j~I(\vec{v}_i,\, \vec{v}_j)
\end{equation}
yields
\begin{equation}
\langle f_\beta|f_\alpha\rangle = \left(\frac{\lambda}{E_d}\right)^{{\cal{A}}_\beta + {\cal{A}}_\alpha - {\cal{A}}_{\beta\alpha}}
\end{equation}
We will be interested in cases for which $Q_\alpha=Q_\beta$ (and likewise for the total energy and momentum). Then 
\begin{equation}
{\cal{A}}_\beta + {\cal{A}}_\alpha - {\cal{A}}_{\beta\alpha}={\cal{B}}_{\beta\alpha}=-\frac{1}{16\pi^2}~\left(\sum_{ij \in \beta}~+~\sum_{ij \in \alpha}~-~2\sum_{i \in \alpha}\sum_{j \in \beta}\right)~e_i\,e_j~v_{ij}^{-1}~ \ln \left(\frac{1+ v_{ij}}{1-v_{ij}}\right)
\end{equation}
Let us call the $\beta$ particles outgoing and the $\alpha$ particles incoming, and define $\eta_i$ to be $+1$ for all outgoing particles and $-1$ for all incoming particles. Then we may write
\begin{equation}
{\cal{B}}_{\beta\alpha}=-\frac{1}{16\pi^2}~\sum_{ij}~\eta_i\,\eta_j\,e_i\,e_j~v_{ij}^{-1}~ \ln \left(\frac{1+ v_{ij}}{1-v_{ij}}\right)
\end{equation}
where the sums are overall outgoing and incoming particles. Notice that when ${\cal{B}}_{\beta\alpha}$ is positive, and to all orders in the electron charge, the overlap vanishes in the $\lambda \to 0$ limit:
\begin{equation}
\langle f_\beta|f_\alpha\rangle = \left(\frac{\lambda}{E_d}\right)^{{\cal{B}}_{\beta\alpha}}
\end{equation}
%%%%%%%%%%%%%%%%%%%%%%%%%%%%%%%%%%%%%%%%%%%%%%%%%%%%%%%%%%%%%%%%%%%%%%%%%%%%%%%%%%%
% \section{Bloch-Norddsieck rate}\label{A3}

%%%%%%%%%%%%%%%%%%%%%%%%%%%%%%%%%%%%%%%%%%%%%%%%%%%%%%%%%%%%%%%%%%%%%%%%%%%%%%%%%%

\end{document}